\documentclass[epj]{svjour}
\usepackage{graphics}
\usepackage{graphicx}
\usepackage{amsmath}
\usepackage{amssymb}
\usepackage{newtxtext}
\usepackage{newtxmath}
\RequirePackage[numbers,sort&compress]{natbib}
\RequirePackage[colorlinks,citecolor=blue,urlcolor=blue,linkcolor=blue]{hyperref}
\makeatletter
\renewcommand{\sffamily}{\rmfamily} 

\makeatother
\hypersetup{
    colorlinks=true,
    linkcolor=blue,
    citecolor=blue,
    urlcolor=blue
}

\usepackage[switch]{lineno}
\journalname{Eur. Phys. J.}

\begin{document}
\begin{sloppypar}

\title{Development of a dual-phase xenon time projection chamber prototype for the RELICS experiment}

\author{Lingfeng Xie\inst{1} \and Jiajun Liu\inst{2} \and Yifei Zhao\inst{1} \and Chang Cai\inst{1} \and Guocai Chen\inst{3} \and Jiangyu Chen\inst{4} \and Huayu Dai\inst{5} \and Rundong Fang\inst{6} \and Hongrui Gao\inst{1} \and Fei Gao \inst{1} \thanks{\emph{Corresponding author: } feigao@tsinghua.edu.cn} \and Jingfan Gu\inst{1} \and Xiaoran Guo\inst{7,8} \and Jiheng Guo\inst{6} \and Chengjie Jia\inst{1} \thanks{Now at: Department of Physics, Stanford University, Stanford, CA 94305, USA} \and Gaojun Jin\inst{3} \and Fali Ju\inst{3} \and Yanzhou Hao\inst{1} \and Xu Han\inst{1} \and Yang Lei\inst{1} \and Kaihang Li\inst{1} \and Meng Li\inst{3} \and Minhua Li\inst{3} \and Ruize Li\inst{9} \and Shengchao Li\inst{9} \and Siyin Li\inst{9} \and Tao Li\inst{3} \and Qing Lin\inst{7,8} \and Sheng Lv\inst{3} \and Guang Luo\inst{2} \and Yuanyuan Ren\inst{5} \and Chuanping Shen\inst{3} \and Mingzhuo Song\inst{1} \and Lijun Tong\inst{7,8} \and Yuhuang Wan\inst{1} \and Xiaoyu Wang\inst{9} \and Wei Wang\inst{2,4} \and Xiaoping Wang\inst{6,10} \and Zihu Wang\inst{3} \and Yuehuan Wei\inst{4} \and Liming Weng\inst{3} \and Xiang Xiao\inst{2} \thanks{\emph{Corresponding author: } xiaox93@mail.sysu.edu.cn} \and Yikai Xu\inst{1} \and Jijun Yang\inst{9} \and Litao Yang\inst{11} \and Long Yang\inst{3} \and Jingqiang Ye\inst{5} \and Jiachen Yu\inst{7,8} \and Qian Yue\inst{11} \and Yuyong Yue\inst{4} \and Tianyuan Zha\inst{5} \and Bingwei Zhang\inst{3} \and Yuming Zhang\inst{1} \and Chenhui Zhu\inst{8} \ (RELICS Collaboration) \thanks{\emph{Collaboration email: } relics@tsinghua.edu.cn} \and Xunan Guo \inst{6} \and Xiaopeng Zhou\inst{6}}

\institute{
Department of Physics \& Center for High Energy Physics, Tsinghua University, Beijing 100084, China
\and
School of Physics, Sun Yat-sen University, Guangzhou 510275, China
\and
CNNC Sanmen Nuclear Power Company, Zhejiang 317112, China
\and
Sino-French Institute of Nuclear Engineering and Technology, Sun Yat-sen University, Zhuhai 519082, China
\and
School of Science and Engineering, The Chinese University of Hong Kong (Shenzhen), Shenzhen, Guangdong 518172, China
\and
School of Physics, Beihang University, Beijing 100083, China
\and
State Key Laboratory of Particle Detection and Electronics, University of Science and Technology of China, Hefei 230026, China
\and
Department of Modern Physics, University of Science and Technology of China, Hefei 230026, China
\and
School of Science, Westlake University, Hangzhou 310030, China
\and
Beijing Key Laboratory of Advanced Nuclear Materials and Physics, Beihang University, Beijing 100191, China
\and
Key Laboratory of Particle and Radiation Imaging (Ministry of Education) \& Department of Engineering Physics, Tsinghua University, Beijing 100084, China}

\date{Received: date / Accepted: date}

\maketitle

\noindent 
\textbf{Abstract}
\enspace
The RELICS (REactor neutrino LIquid xenon Coherent elastic Scattering) experiment aims to detect coherent elastic neutrino-nucleus scattering from reactor antineutrinos using a dual-phase xenon time projection chamber (TPC). 
To validate the detector concept and ensure technical reliability for the full-scale experiment, a dedicated prototype was designed, constructed, and operated. 
This work presents an overview of the design, construction, and operational performance of the prototype, with emphasis on its major subsystems, including the TPC, cryogenics and xenon purification systems, slow control, and data acquisition. 
During operation, the detector demonstrated the capability to achieve a sub-keV energy threshold required for the RELICS physics program, as reflected by a measured single electron gain of (34.30~$\pm$~0.01~(stat.))~PE/e$^-$ and the successful detection of $\mathrm{0.27~keV}$ L-shell decay events from $\mathrm{^{37}Ar}$. 
In addition, essential data analysis techniques and simulation frameworks were developed and validated, establishing the technical foundation for future RELICS operations. 
The successful construction and operation of this prototype confirm the feasibility of the core technologies and provide the experimental basis for the full-scale RELICS detector.
\par\vspace{2em}

\section{Introduction}
\label{sec:intro}

Coherent elastic neutrino-nucleus scattering (CE$\mathrm{\nu}$NS) was first predicted in 1974 by Freedman within the Standard Model (SM) framework~\cite{PhysRevD.9.1389}. 
It describes a process in which a neutrino scatters coherently off an entire nucleus at sufficiently low momentum transfer, with the scattering amplitudes of individual nucleons combining coherently to enhance the total amplitude.
This results in a cross section scaling approximately with $\mathrm{N^2}$ (neutron number squared).
However, direct observation remained elusive for over four decades due to the extraordinarily low nuclear recoil energies (keV or sub-keV range) and formidable backgrounds from environmental, material radioactivity, as well as the intrinsic backgrounds of the detector.

CE$\mathrm{\nu}$NS measurement is important for both particle physics and astrophysics.
As a neutral current process, it provides a high precision probe of the SM electroweak sector.
Deviations from theoretical predictions could indicate beyond SM physics, such as non-standard neutrino interactions (NSIs), light sterile neutrinos, or anomalous neutrino magnetic moments~\cite{Coloma:2017,PhysRevC.86.024612,Chattaraj:2025}. 
CE$\mathrm{\nu}$NS also governs energy transport in supernova core collapse dynamics~\cite{PhysRevC.95.025801}.

The experimental breakthrough came in 2017 when COHERENT first observed CE$\mathrm{\nu}$NS using cesium iodide detector and neutrinos from the Spallation Neutron Source (SNS)~\cite{doi:10.1126/science.aao0990}, followed by confirmations with liquid argon and germanium detectors~\cite{PhysRevLett.126.012002, PhysRevLett.134.231801}. 
The SNS provides high flux, pulsed neutrinos with tens of MeV energies.
In 2025, CONUS+ achieved the direct observation of reactor antineutrino CE$\mathrm{\nu}$NS using germanium detectors, where the $<$10~MeV spectrum produces smaller nuclear recoils, advancing low-threshold technology~\cite{Ackermann:2025}. 
Large liquid xenon (LXe) experiments (XENONnT, PandaX-4T, LZ) now search for CE$\mathrm{\nu}$NS from solar neutrinos, studying the neutrino fog background for Weakly Interacting Massive Particle (WIMP) searches~\cite{PhysRevLett.133.191002,PhysRevLett.133.191001,LZ:2025igz}.
The RED-100 experiment demonstrated the feasibility of two-phase xenon time projection chamber technology for CEvNS searches at a reactor site and placed the limit for reactor neutrino CEvNS on xenon nuclei~\cite{PhysRevD.111.072012}.

The RELICS (REactor neutrino LIquid xenon Coherent elastic Scattering) experiment~\cite{cai2024relicsreactorneutrinoliquid} aims to perform a precise measurement of CE$\mathrm{\nu}$NS from reactor antineutrinos using a dual-phase xenon TPC.
Located approximately 25 meters from a reactor core at the Sanmen Nuclear Power Plant, the 50 kg RELICS detector will be exposed to a large antineutrino flux of roughly $10^{13}~\bar{\nu}_\mathrm{e}\,\mathrm{cm}^{-2}\,\mathrm{s}^{-1}$~\cite{Qian:2018wid}, with a target of detecting approximately 5,000 CE$\mathrm{\nu}$NS events annually.  
The choice of dual-phase xenon TPC technology is motivated by its demonstrated success in leading particle physics experiments~\cite{Aprile:2009dv}. 
LXe has high yields of scintillation light and ionization electrons.
It contains no long-lived radioactive isotopes and has a relatively high liquefaction temperature, which facilitates stable cryogenic operation.
Particle interactions in a dual-phase xenon TPC generate two signals. 
The primary scintillation signal (S1) arises from the de-excitation of xenon excimers and from electron-ion recombination in the liquid phase. 
The secondary signal (S2) is produced by electroluminescence when ionization electrons are extracted into the gaseous xenon phase.
However, detecting CE$\mathrm{\nu}$NS utilizing a dual-phase xenon TPC presents distinct challenges. 
First, the nuclear recoils induced by reactor neutrinos are extremely low energy, typically on the order of keV.
Secondly, while extensive shielding mitigates backgrounds from cosmic-ray-induced and environmental radioactivity, the sensitivity of the experiment is primarily limited by delayed electron (DE) emissions~\cite{PhysRevD.106.022001}. 
In the low-energy regime relevant to CE$\mathrm{\nu}$NS, these DE emissions are expected to be the dominant background.
The physical origin of DEs is not yet fully understood.
They are generally attributed to photoionization of impurities or metallic surfaces shortly after large S2 signals, and to long timescale releases associated with impurities or field effects in LXe.
High energy depositions from cosmic muons can further contribute by generating few-electron S2 tails that overlap with genuine low energy events.
These DE backgrounds remain a major limitation to achieving low-background CE$\mathrm{\nu}$NS sensitivity.
These challenges necessitate a detector with an exceptionally low energy threshold and powerful background discrimination capabilities. 
To achieve the low energy threshold, the RELICS experiment will employ an S2-only analysis strategy~\cite{PhysRevD.111.062006}.
This approach relies exclusively on the S2 generated via electroluminescence, as the S1 is too faint to be detected.

To validate the design concept and improve technical solutions, a dedicated prototype has been constructed and operated.
In this paper, we describe the design, construction, and operation of a dual-phase xenon TPC prototype developed to validate the overall design and the key technologies for RELICS. 
The prototype TPC is detailed in Sec.~\ref{sec:prototype}, followed by the descriptions of the cryogenics, circulation and purification system in Sec.~\ref{sec:cooling_purification}, the slow control (SC) system in Sec.~\ref{sec:slow_control}, and the data acquisition system in Sec.~\ref{sec:daq}.
The analysis of calibration data is presented in Sec.~\ref{sec:analysis}, and we conclude in Sec.~\ref{sec:conclusion}.

\section{Design and Construction of the Prototype TPC}
\label{sec:prototype}

\subsection{Prototype TPC Design and Instrumentation}

\begin{figure*}[!htbp]
\includegraphics[width=\linewidth]{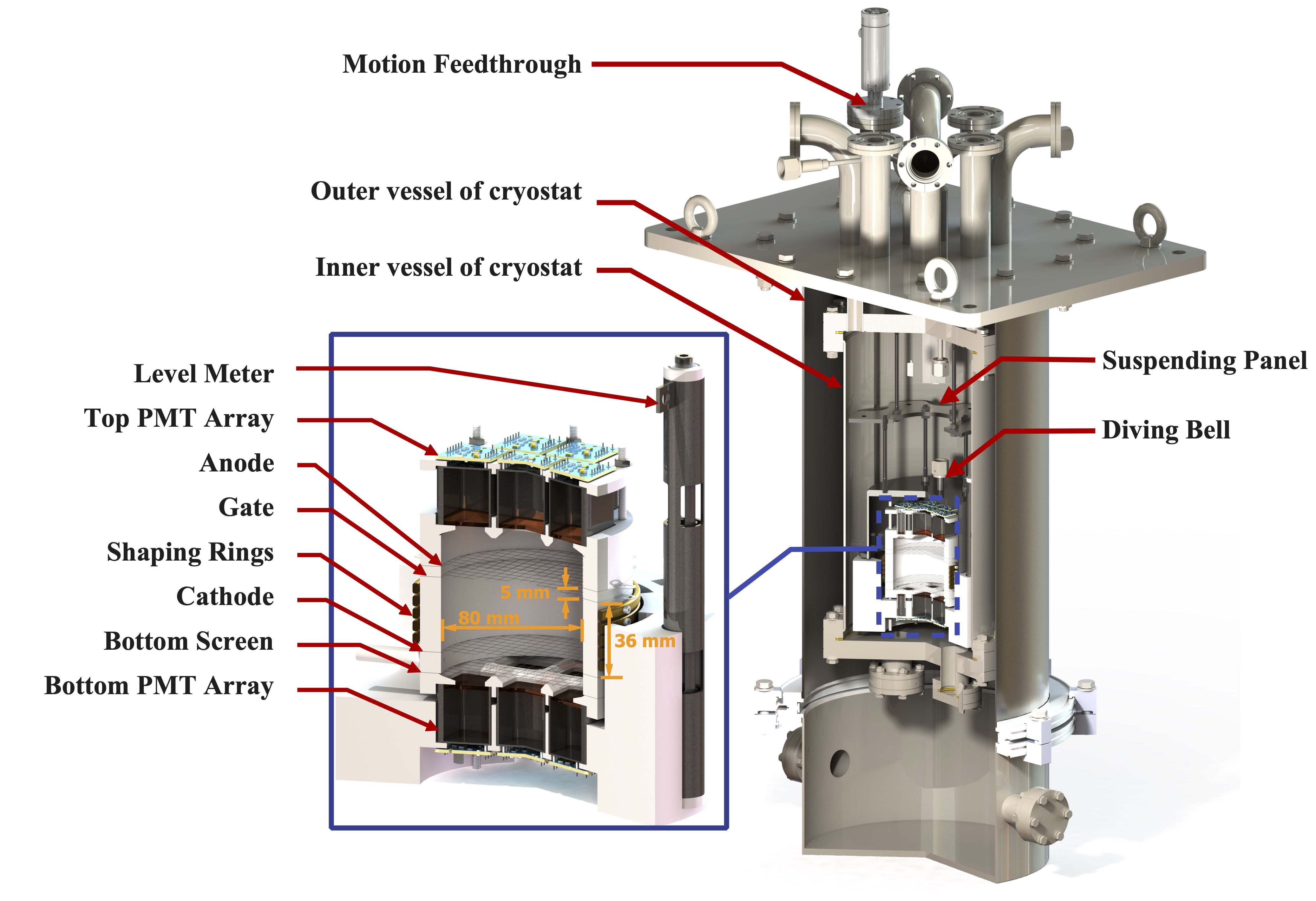}
    \caption{ Schematic cross-section of the dual-phase xenon prototype.
    The main cutaway view shows the cryostat vessels and support structures, while the enlarged inset highlights the TPC layout. 
    The top and bottom photomultiplier (PMT) arrays detect scintillation and electroluminescence photons. 
    All inner components are suspended from the top flange by the structural suspension panel.
    Components including the PMT arrays, field cage, diving bell and level meters are labeled.
    }
    \label{fig:prototype}
\end{figure*}

To validate the design of the RELICS experiment, a dual-phase xenon TPC prototype was developed.
The design of the prototype is shown in Fig.~\ref{fig:prototype}.
The TPC features a cylindrical inner volume with a diameter of $\mathrm{80~mm}$ and a height of $\mathrm{36~mm}$, corresponding to an active liquid xenon mass of approximately $\mathrm{0.55~kg}$.
This inner region is enclosed by Polytetrafluoroethylene (PTFE) walls, chosen for their high reflectivity for vacuum ultraviolet (VUV) scintillation light at $\mathrm{178~nm}$ emitted by xenon~\cite{Kravitz:2020}. 
The entire TPC assembly is suspended via stainless steel screw rods from the top flange of a double-walled, vacuum-insulated cryostat, which ensures the thermal isolation required for stable cryogenic operation.

The photon detection system consists of two arrays of seven \textit{Hamamatsu R8520-406} 1-inch photomultiplier tubes (PMTs), arranged in a hexagonal pattern at the top and bottom of the TPC, totaling 14 detection channels. 
The PMT model is the same as the one selected for the full-scale RELICS detector.
To calibrate PMT gains, an internal LED and an optical fiber delivering light from an external LED are installed. 
The optical fiber is utilized to avoid electromagnetic interference during PMT calibration.
Conversely, the internal LED provides higher light intensity for cases where the fiber output is insufficient, though it introduces some electromagnetic interference.
Precise control of the operational parameters is critical for stable performance.
The LXe level, which directly impacts the single electron gain (SEG), is maintained by a diving bell with a mechanically adjustable level-setting mechanism.
A gas inlet line pressurizes the bell interior to regulate the liquid level.
Simultaneously, a welded bellows outlet tube, connected to a linear motion feedthrough, extends into the diving bell to exhaust excess gas.
The vertical position of this tube is adjustable from outside the cryostat via the motion feedthrough.
As the gas pressure pushes the liquid down, excess gas escapes through the outlet tube once the interface reaches the outlet tube bottom opening.
Based on the principle of communicating vessels, the liquid level inside the bell equilibrates with the height of the outlet tube bottom opening.
The level is actively monitored by capacitive liquid level meters~\cite{llm}.
Furthermore, to ensure safe and stable operation, the thermodynamic state of the detector is continuously monitored by five temperature sensors and two pressure transmitters.
These sensors are placed to track thermal and pressure conditions.
An alert is automatically triggered by the SC system (detailed in Sec.~\ref{sec:slow_control}) if the pressure exceeds safe operational limits, initiating predefined emergency procedures.

\subsection{Electric field and field cage design}
\label{subsec:field_cage}

The detection of ionization electrons produced in LXe relies on two electric fields: a drift field that transports electrons vertically through the liquid, and a stronger extraction field at the liquid-gas interface that pulls electrons into the gas phase, where the S2 signal is generated. 
These fields are established by the field cage of the TPC, which consists of a series of electrodes and field shaping rings.
The electrodes include the anode, the gate, the cathode, and the bottom screen.
All electrodes are constructed from chemically etched stainless-steel meshes with a wire pitch of $\mathrm{5~mm}$ and a wire diameter of $\mathrm{0.1~mm}$. 
Between these electrodes, a cylindrical PTFE insulator serves as both structural support and reflector, featuring an inner diameter of $\mathrm{8~cm}$ and an outer diameter of $\mathrm{10~cm}$.
To enhance field uniformity within the drift region, five field shaping rings are concentrically mounted along the outer surface of the PTFE cylinder between gate and cathode, spaced evenly along the drift length and connected by a resistive voltage divider chain (\textit{Ohmite MOX200001007FE}, with a resistance of $\mathrm{1~G\Omega}$).
Each mesh undergoes surface cleaning, acid etching, and passivation to reduce electron emission from microscopic imperfections under high-field conditions~\cite{Aprile2022}.

The anode was originally designed to operate at ground potential to suppress gas events. 
During the operation, the maximum stable cathode voltage was found to be around $\mathrm{-3.2~kV}$, beyond which electrical breakdown occurred. 
To achieve adequate SEG and electron extraction efficiency (EEE), a positive high voltage was applied to the anode. 
The gate and anode are biased at $\mathrm{-2600~V}$ and $\mathrm{+1600~V}$ respectively, establishing the extraction and electroluminescence field across a $\mathrm{5~mm}$ gap. 
This extraction field is approximately $\mathrm{4~kV/cm}$ in the liquid. 
The top PMTs operate stably without degradations of performance under this configuration.
The cathode is held at $\mathrm{-3248~V}$, establishing a drift field of $\mathrm{\sim 180~V/cm}$ across the $\mathrm{36~mm}$ drift region between the cathode and the gate. 
A bottom screening electrode located $\mathrm{10~mm}$ below the cathode, biased at $\mathrm{-800~V}$, shields the bottom PMT array.

A \textit{CAEN N1470} power supply unit is utilized to provide the required high voltage, ensuring precise voltage control and stable operation.
To transmit the high voltage from the high voltage source to the electrodes, an SHV-10 coaxial feedthrough (\textit{MPF A1601-3-CF}) is employed. 
At the atmospheric side, \textit{RG086} cables facilitate the connection between the high voltage source and the SHV-10 port on the feedthrough. 
Inside the cryostat, Kapton-insulated wires are used to link the vacuum side of the feedthrough to the electrode terminals. 
Conductors are extensively insulated using shrink tubing to minimize the risk of dielectric breakdown.

\begin{figure}[htbp]
	\includegraphics[width=1\linewidth]{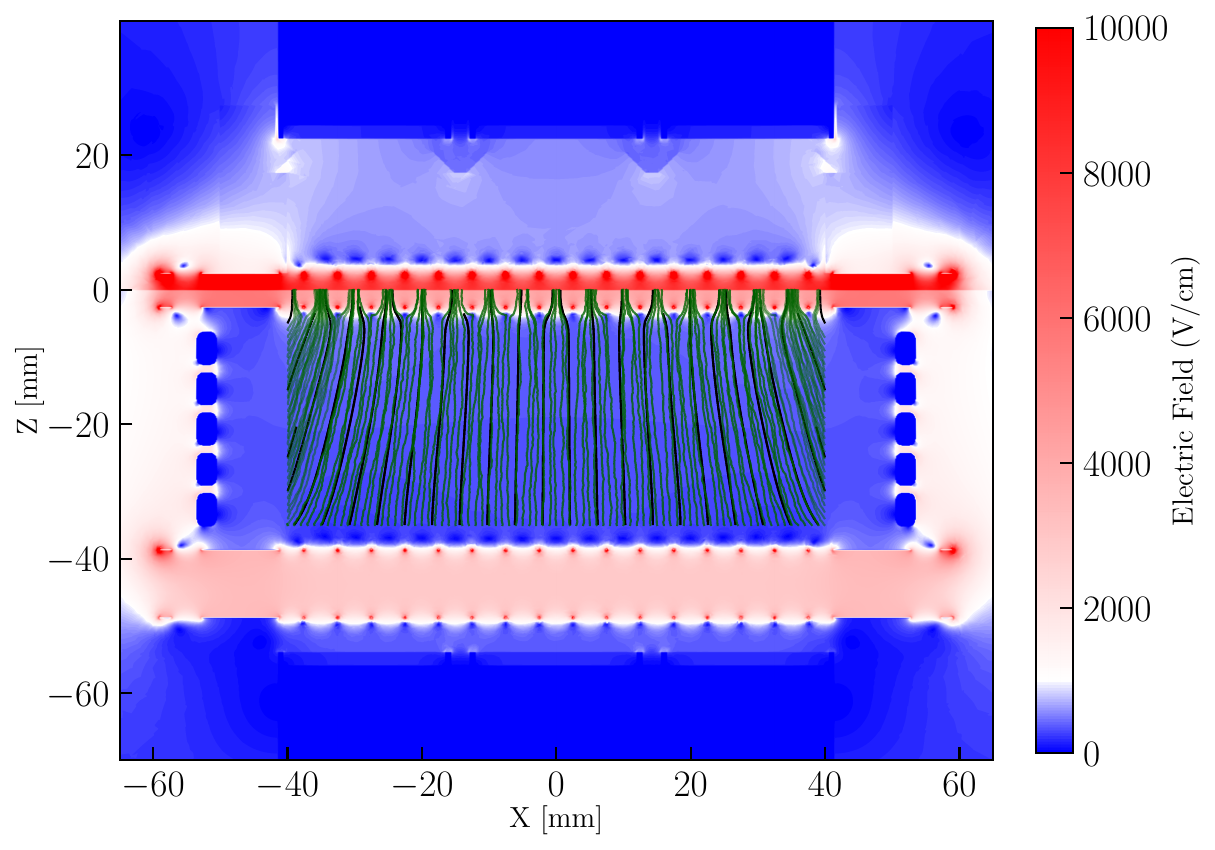} 
	\caption{ The electric field magnitude distribution in the X-Z cross-section within the TPC, simulated using COMSOL Multiphysics. 
    The color scale indicates the field strength in unit of $\mathrm{V/cm}$. 
    The simulation visualizes the high-field extraction region between the gate and anode (top red area), and the drift field in the central volume, maintained by the field-shaping rings. 
    The black lines represent the electric field lines, and the green streamlines trace electron drift paths from various initial positions, demonstrating their successful transport to the liquid-gas interface for extraction.
    }
	\label{fig:comsol_field}
\end{figure}

\begin{figure}
	\includegraphics[width=1\linewidth]{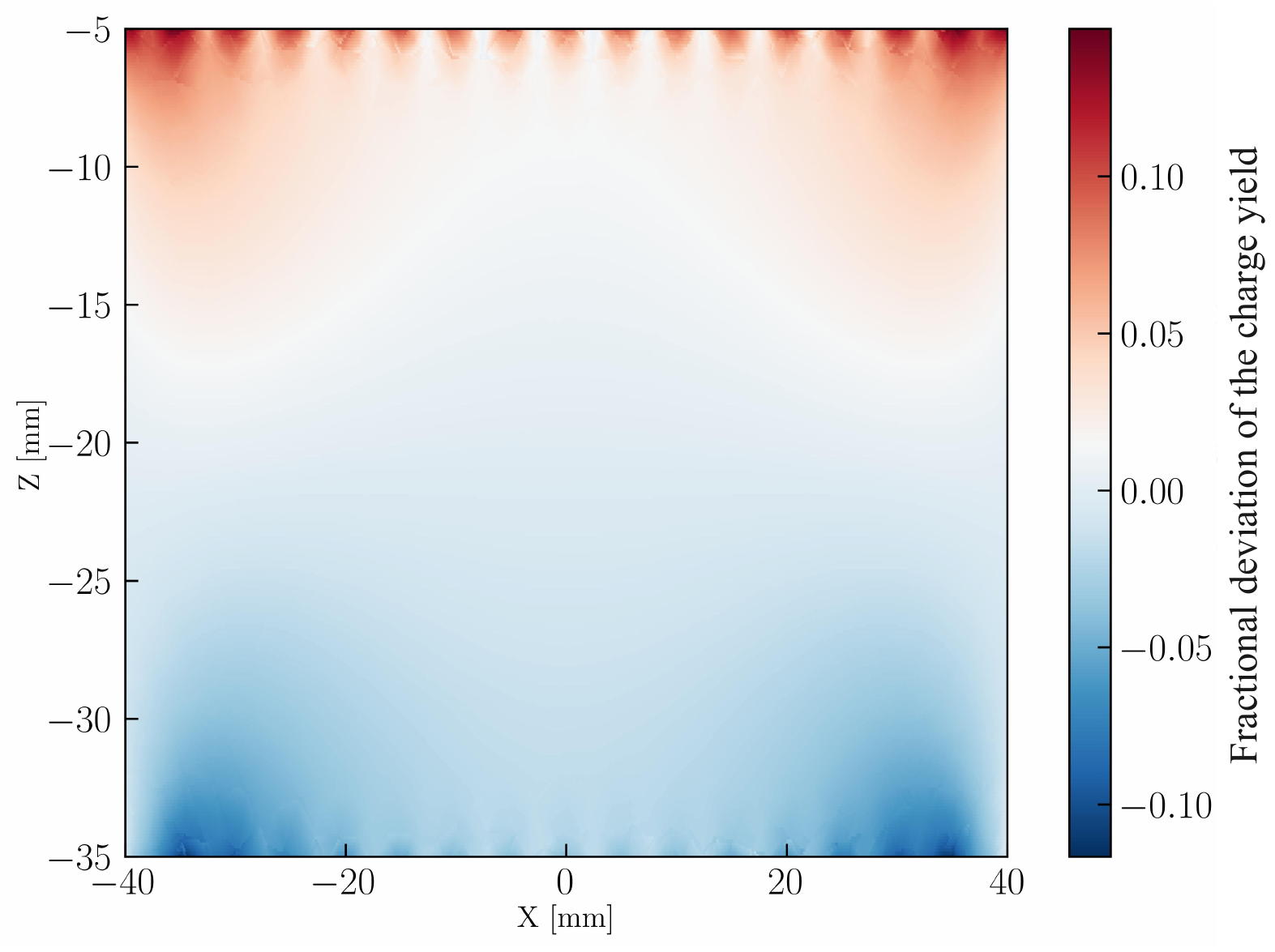}
	\caption{The map shows the fractional deviation of the charge yield (CY) in the X-Z cross-section within the TPC inner drift volume, defined as $\mathrm{\frac{\Delta CY(x,y,z)}{CY_{\mathrm{ave}}} = \frac{CY(x,y,z)-CY_{\mathrm{ave}}}{CY_{\mathrm{ave}}}}$. }
    \label{fig:QY}
\end{figure}

Achieving a uniform drift field is a fundamental design goal for a dual-phase xenon TPC. 
Non-uniform electric fields can distort electron drift trajectories, increasing the probability of charge loss on chamber walls and inducing position-dependent variations in charge and light yields~\cite{10.3389/fdest.2024.1480975}. 
Light and charge yield simulations are required to generate position-dependent correction maps used to correct TPC signals.
The three-dimensional electrostatic simulation by COMSOL Multiphysics~\cite{comsol} revealed the field distribution with the electrodes configuration, as shown in Fig.~\ref{fig:comsol_field}.
Based on this field map, the position-dependent charge yield correction was derived using the NEST model~\cite{10.3389/fdest.2024.1480975}, and the resulting correction map for $\mathrm{^{83m}Kr}$ is presented in Fig.~\ref{fig:QY} as an example.
The extraction field is established between the gate and the anode, with the liquid surface maintained approximately $\mathrm{2.5~mm}$ above the gate mesh.
The extraction field in LXe is simulated to be about $\mathrm{4.3~kV/cm}$, taking into account the electron drift trajectories towards the liquid surface. 
Under this field, the EEE is roughly estimated to be in the range of $70\%$ to $90\%$ based on various experimental measurements~\cite{Xu,Gushchin,XENON100_EEE,LUX_EEE,PIXeY}.

\subsection{Light signal readout and transfer}

Detection of VUV photons from both scintillation and electroluminescence processes is accomplished using arrays of PMTs, located at the top and bottom of the TPC.
These PMTs are arranged in a hexagonal pattern to ensure geometric coverage.
Each PMT has a photon detection efficiency of approximately $\mathrm{30\%}$ for xenon scintillation light at $\mathrm{178~nm}$ and operates at a bias voltage of $\mathrm{-800~V}$.

Each PMT is connected via two independent cables: a 20~AWG single-core wire delivering the high voltage bias from a \textit{CAEN A7030SN} supply, and a 26~AWG, $\mathrm{50~\Omega}$ coaxial cable for analog signal readout.
Both cable types feature polyimide (PI) insulation and silver-plated copper conductors, selected specifically for their low outgassing rates and operation at cryogenic temperatures down to $\mathrm{170~K}$.
Furthermore, the coaxial signal cable provides shielding to suppress noise pickup, with its $\mathrm{50~\Omega}$ characteristic impedance ensuring proper matching with the readout electronics.

The connection between the internal cryogenic cabling and the external readout electronics described in Sec.~\ref{sec:daq} is established through multi-pin feedthroughs mounted on CF35 flanges.
Commercial feedthroughs \textit{MPF A8232-5-CF}, compliant with the \textit{MIL-C-26482} standard, provide 32 gold-plated conductors rated for 1~kV per pin, offering low-resistance, high-integrity electrical connections.
A total of two such feedthroughs are used to accommodate the 14 PMTs, with separate pins assigned for the signal, ground, and high voltage supply of each PMT channel.

Accurate reconstruction of photoelectron (PE) counts from PMT waveforms requires precise calibration of the individual PMT gain and resolution.
The calibration is performed by illuminating the PMTs with controlled LED pulses driven by a signal generator.
The resulting charge spectra are recorded and statistically analyzed to obtain the single-photoelectron (SPE) response.

The SPE spectrum is modeled as a convolution of Poisson-distributed PE statistics with the Gaussian response of the PMT amplification process.
The overall probability density function  describing the collected charge distribution can be expressed as:

\begin{flalign}
\label{eq:pmt_calibration}
\mathrm{f(x)} &= \mathcal{N}_{\mathrm{ped}}\!\left( \mathrm{x}; \mu_{\mathrm{ped}}, \sigma_{\mathrm{ped}} \right)  +  \nonumber \\
&\sum_{\mathrm{n=1}}^{\mathrm{k}} \mathcal{P}_{\mathrm{pe}}\!\left( \mathrm{n}; \lambda_{pe} \right)
\cdot \mathcal{N}_{pe}\!\left( \mathrm{x}; \mathrm{nG}, \mathrm{\sqrt{n\sigma_{G}^{2} + \sigma_{\mathrm{ped}}^{2}}} \right) ,
\end{flalign}

where $\mathrm{x}$ is the measured charge in units of electron count.
The term $\mathrm{\mathcal{N}_{\mathrm{ped}}}$ describes the pedestal distribution due to baseline fluctuations, characterized by the electronic noise standard deviation $\sigma_{\mathrm{ped}}$. 
$\mathrm{\mathcal{P}_{pe}}$ denotes the Poisson probability of detecting $\mathrm{n}$ PE with mean $\mathrm{\lambda_{pe}}$.
$\mathrm{\mathcal{N}_{pe}}$ represents the Gaussian probability of the collected electron distribution of the response to SPE.
$\mathrm{G}$ is the mean single PE gain, and $\mathrm{\sigma_{G}}$ represents the intrinsic gain fluctuation of the electron multiplication process.
$\mathrm{\sigma_{\mathrm{ped}}}$ also accounts for the electronic noise contribution to the resolution of the PE peaks (the second term in the square root).

The raw PMT waveforms are recorded in analog-to-digital converter (ADC) counts. 
The integrated pulse area $\mathrm{A_{\text{ADC}}}$ (in units of ADC $\cdot$ samples) is converted to the number of electrons $\mathrm{N_e}$ (variable $\mathrm{x}$ in Eq.~\ref{eq:pmt_calibration}) using the following relation:
\begin{equation}
\mathrm{ N_e = A_{\text{ADC}} \cdot \frac{V_{\text{range}}}{2^{14}} \cdot \frac{t_s}{R \cdot q_e} }
\end{equation}
where $\mathrm{V_{\text{range}} = 0.5~V}$ is the digitizer dynamic range, $\mathrm{2^{14}}$ represents the ADC resolution of the digitizers, $\mathrm{t_s = 4~ns}$ is the sampling period, $\mathrm{R = 50~\Omega}$ is the input impedance, and $\mathrm{q_e}$ is the elementary charge. 
This yields a conversion factor of approximately $\mathrm{1.52 \times 10^4}$ electrons per unit $\mathrm{A_{\text{ADC}}}$.

\begin{figure}[!htbp]
    \centering
    \includegraphics[width=\linewidth]{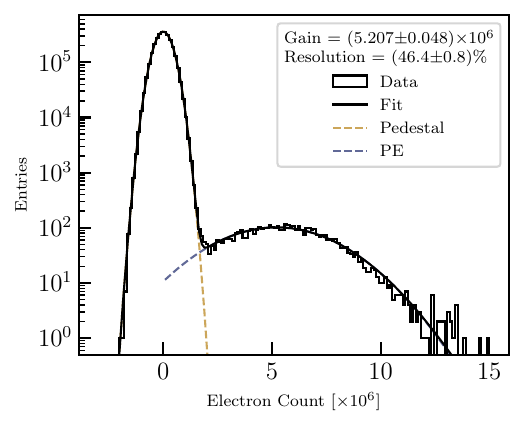}
    \caption{
    Representative single-photoelectron charge spectrum from a PMT Channel.
    The black histogram shows experimental data, fitted with a model (dashed curve) comprising a pedestal (yellow) and photoelectron  component (blue).
    The fit yields a gain of $\mathrm{5.21 \times 10^6}$ at $\mathrm{-800~V}$.}
    \label{fig:pmt_calibration}
\end{figure}

Figure~\ref{fig:pmt_calibration} shows a representative SPE charge spectrum obtained from the calibration of a PMT channel.
The fitted model matches the measured distribution, yielding a mean gain of $\mathrm{G = (5.207 \pm 0.048) \times 10^{6}}$ for an operating bias of $\mathrm{-800~V}$, demonstrating the capability to resolve individual photoelectrons and enabling low-threshold signal detection.
This calibration result serves as a reference for determining the number of detected photons in subsequent analysis.

\section{Xenon handling system}
\label{sec:cooling_purification}

The operation of a dual-phase xenon TPC requires maintaining the xenon in a cryogenic liquid state while ensuring high purity to achieve long electron drift lifetime.
To fulfill these two essential conditions, the RELICS prototype is equipped with an integrated cryogenics and purification system.

The cryogenic subsystem provides the cooling power necessary to liquefy xenon near its operational temperature of approximately 170~K. 
A closed-loop refrigeration circuit, based on a Gifford-McMahon (GM) cooler, maintains a steady liquid-gas equilibrium inside the cryostat, enabling long-term detector stability.

Additionally, a continuous xenon circulation and purification loop removes electronegative impurities such as O$_2$ and H$_2$O that could otherwise capture drifting electrons. 
LXe is evaporated, purified in the gas phase through a heated getter, and recondensed into the detector. 
The flow is controlled through a series of valves and stainless-steel piping, which regulate circulation paths between the detector, purifier, and cryogenics.

This supporting system ensures a thermally stable and high purity xenon environment, which is critical for stable detector operation and sustaining efficient charge transport.

\subsection{Cryostat design}
\label{subsec:cryostat}

The prototype employs a custom-designed cryostat engineered to maintain LXe at its operational temperature of approximately 170~K.
As shown in Fig.~\ref{fig:prototype}, the cryostat consists of a cylindrical, double-walled stainless-steel vessel supported by a rigid aluminum profile frame.
The frame, constructed from extruded aluminum profiles, is mounted on industrial-grade casters that enable easy repositioning within the laboratory. 
Each caster is equipped with an integrated, individually adjustable leveling foot, ensuring both mobility and stable positioning.

The cryostat comprises two main components. 
The inner vessel houses the TPC and is sealed at the top with a DN200CF flange, where this ultra-high vacuum standard ensures a reliable metal-to-metal hermetic seal. 
The outer vessel serves as the vacuum jacket and is closed with a larger ISO-K 300 flange to provide the necessary vacuum enclosure for thermal insulation.
Thermal insulation is achieved primarily through a high vacuum between the inner and outer vessel walls, effectively suppressing conductive and convective heat transfer from the surrounding environment. 
To further minimize radiative heat transfer, the vacuum gap is lined with a multi-layer insulation system composed of alternating layers of fiberglass paper and aluminum foil.
This combination of high vacuum and foil provides thermal isolation, protecting the cryostat interior from external heat loads.

Eight CF35 ports are arranged on the top of the cryostat to provide interfaces for the various feedthroughs and piping required by the system. 
These ports host multi-pin electrical feedthroughs for PMT signal readout and high voltage delivery, SMA-A coaxial feedthroughs for the capacitive level meters, a welded-bellows linear-motion feedthrough for adjusting the diving-bell height, and a pair of vacuum-insulated transfer lines for circulating LXe into and out of the detector.

This cryostat design offers a platform for constructing TPC, ensuring stable operation, efficient thermal management, and operational safety.

\subsection{Cryogenic and liquefaction system}
\label{subsec:cooling}

The continuous operation of the detector relies on a refrigeration system to liquefy the purified gaseous xenon (GXe) and to manage the total thermal load from the environment and the recirculation process~\cite{Zhao_2021, Aprile_2012}.
The core of this system is a \textit{PengLi KDE400SA} series GM cryocooler, which provides approximately 180~W of cooling power at its operational temperature.
The GM cycle is a well-established cryogenic technology that uses a closed loop of high-pressure helium gas, driven by a \textit{KDC6000V} compressor. 
The gas undergoes periodic, adiabatic expansion at the cold head, efficiently extracting thermal energy.
The cold head of cryocooler is thermally coupled to a condenser inside the liquefier, where it chills the incoming GXe below its boiling point.

\begin{figure}[htbp]
    \includegraphics[width=1\linewidth]{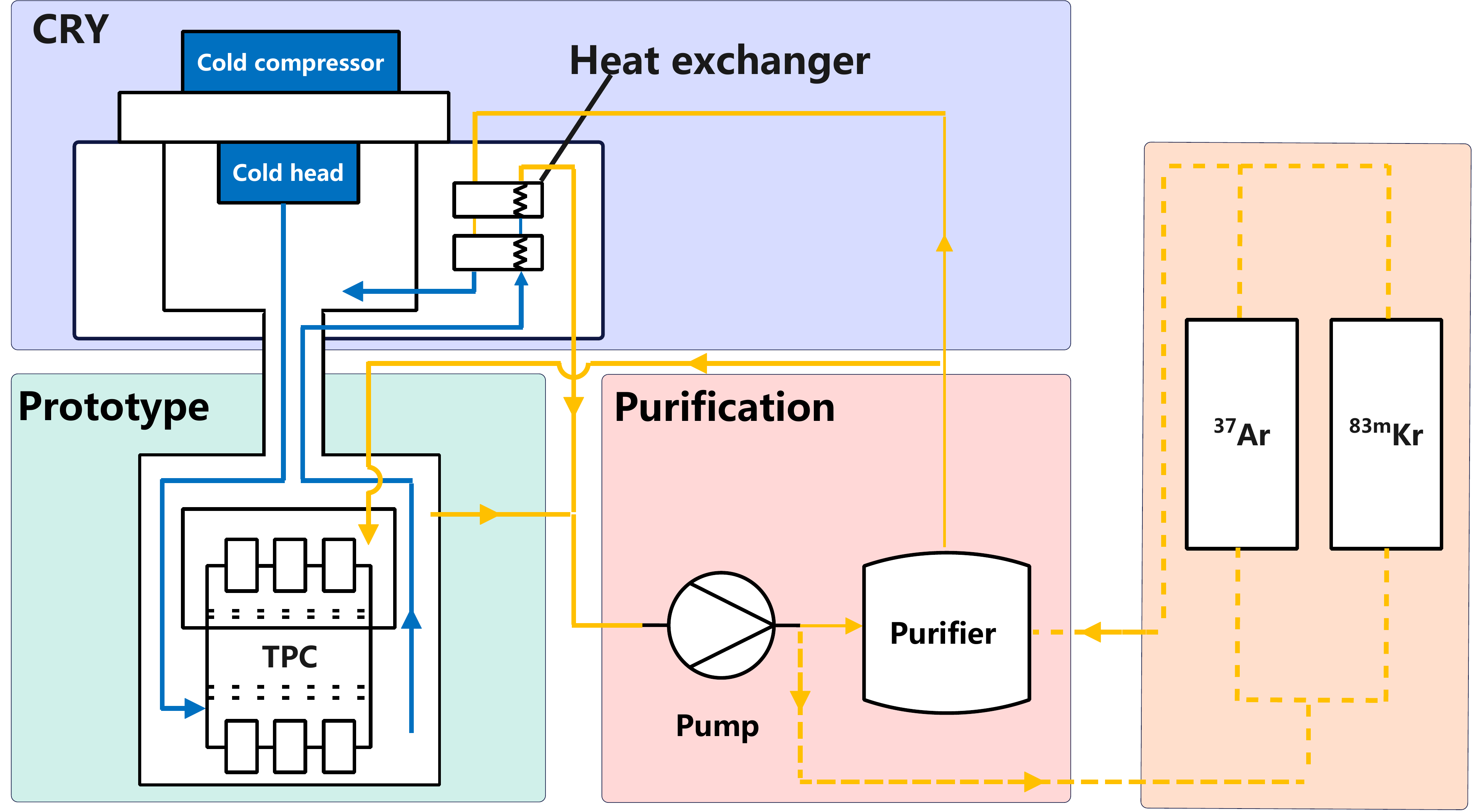}
    \caption{Schematic diagram of the RELICS prototype cryogenic system and xenon handling infrastructure. Blue lines denote the LXe flow path, orange lines denote the GXe circulation used for purification, and orange dashed lines denote the path of the calibration source injection for $\mathrm{^{37}Ar}$ and $\mathrm{^{83m}Kr}$.}
    \label{fig:control_panel}
\end{figure}

To optimize thermal efficiency and minimize the heat load on the cryocooler, the system incorporates two counter-flow plate heat exchangers arranged in series. 
As shown in Fig.~\ref{fig:control_panel}, room-temperature GXe returning from the purification loop undergoes a two-stage cooling process: it is first pre-cooled in a primary (larger) heat exchanger by the cold GXe exiting the detector vessel, followed by further cooling in a secondary (smaller) heat exchanger. 
This cascaded heat exchange brings the GXe close to its condensation point before it reaches the cold head for final liquefaction.
A portion of the cooled xenon is liquefied and flows into the TPC through a phase separator, while the remaining GXe is further cooled by the cold head to ensure complete condensation.
This strategy effectively recovers and reuses the cooling capacity of the cold exhaust gas, significantly enhancing the system overall thermal efficiency.

The cooling output of the GM cooler is not easily modulated and typically provides more cooling capacity than required for the system. 
Therefore, precise temperature regulation is achieved through active thermal balancing using resistive heaters to counteract the excess cooling power.
Two high-power resistive heaters, each with a resistance of $\mathrm{72~\Omega}$ and a maximum thermal output of $\mathrm{300~W}$, are installed on the cold head.
These heaters are controlled by a feedback loop system that dynamically adjusts the heater power in real-time to compensate for thermal fluctuations and maintain stable thermal equilibrium within the detector.
This active temperature control strategy ensures optimal operating conditions, and the detailed implementation of the Proportional-integral-derivative (PID) control algorithm will be discussed comprehensively in subSec.~\ref{subsec:PID}.

\subsection{Xenon recirculation and purification system}
\label{subsec:circulation_purification}

A closed-loop circulation and purification system is essential for managing gas flow and maintaining the ultra-high purity of the xenon.
As shown in the schematic in Fig.~\ref{fig:control_panel}, the system is built around a primary circulation loop that continuously draws LXe from the detector and returns it to the liquefier.
The driving force for this circulation is provided by a metal bellows pump (\textit{Senior Flexonics MB-602}), chosen for its gas-tightness and oil-free operation, which prevents contamination.
The single-phase motor of the pump is driven by a high-performance \textit{EURA E2000} series variable-frequency drive (VFD), which has been demonstrated to reach up to 15~SLPM for GXe at room temperature, providing a valuable performance baseline for the full-scale RELICS experiment.

Purity is maintained by a \textit{Simpure 9NG} hot getter integrated into the primary loop.
This unit employs a chemical reaction at high temperatures to remove electronegative impurities such as O$_2$ and H$_2$O down to part-per-billion (ppb) level, which is critical for achieving long electron lifetimes, which will be defined and detailed in subSec.~\ref{subsec:lifetime}.
For gas handling and inventory management, the system includes two high-pressure cylinders for xenon storage, typically maintained at approximately 60~bar, and a liquid nitrogen (LN$_2$) dewar for xenon recovery.
Xenon can be injected into the system in a controlled manner via a pressure-reducing valve and mass flow controller (MFC).

For detector calibration, an auxiliary branch is connected to the main loop, as depicted in Fig.~\ref{fig:control_panel}. 
This branch is equipped with a dedicated port for connecting gaseous calibration sources, such as $\mathrm{^{37}Ar}$ and $\mathrm{^{83m}Kr}$. 
The source container is first connected to a small, intermediate dilution volume. 
This volume is evacuated before the source is introduced, allowing for the precise, controlled injection of a known small quantity of the calibration gas into the main xenon circulation stream.

\section{The slow control system}
\label{sec:slow_control}

The reliable and safe operation of the RELICS prototype detector relies on the real-time monitoring and regulation of integrated subsystems.
A semi-distributed SC system has been developed to fulfill this role, providing continuous, real-time oversight of key operational parameters such as temperature, pressure, flow rate and LXe level.

\subsection{Slow control system architecture}

The architecture of this SC system is designed with inherent modularity and scalability, ensuring its direct applicability to the full-scale RELICS experiment.
The hardware interfaces, software infrastructure, and control logic can be readily expanded to accommodate the larger-scale operational requirements and more complex interdependence of the final detector.
This design guarantees a seamless transition from the prototype validation phase to the deployment of a robust, long-term scientific instrument.

\begin{figure*}
    \centering
    \includegraphics[width=0.85\linewidth]{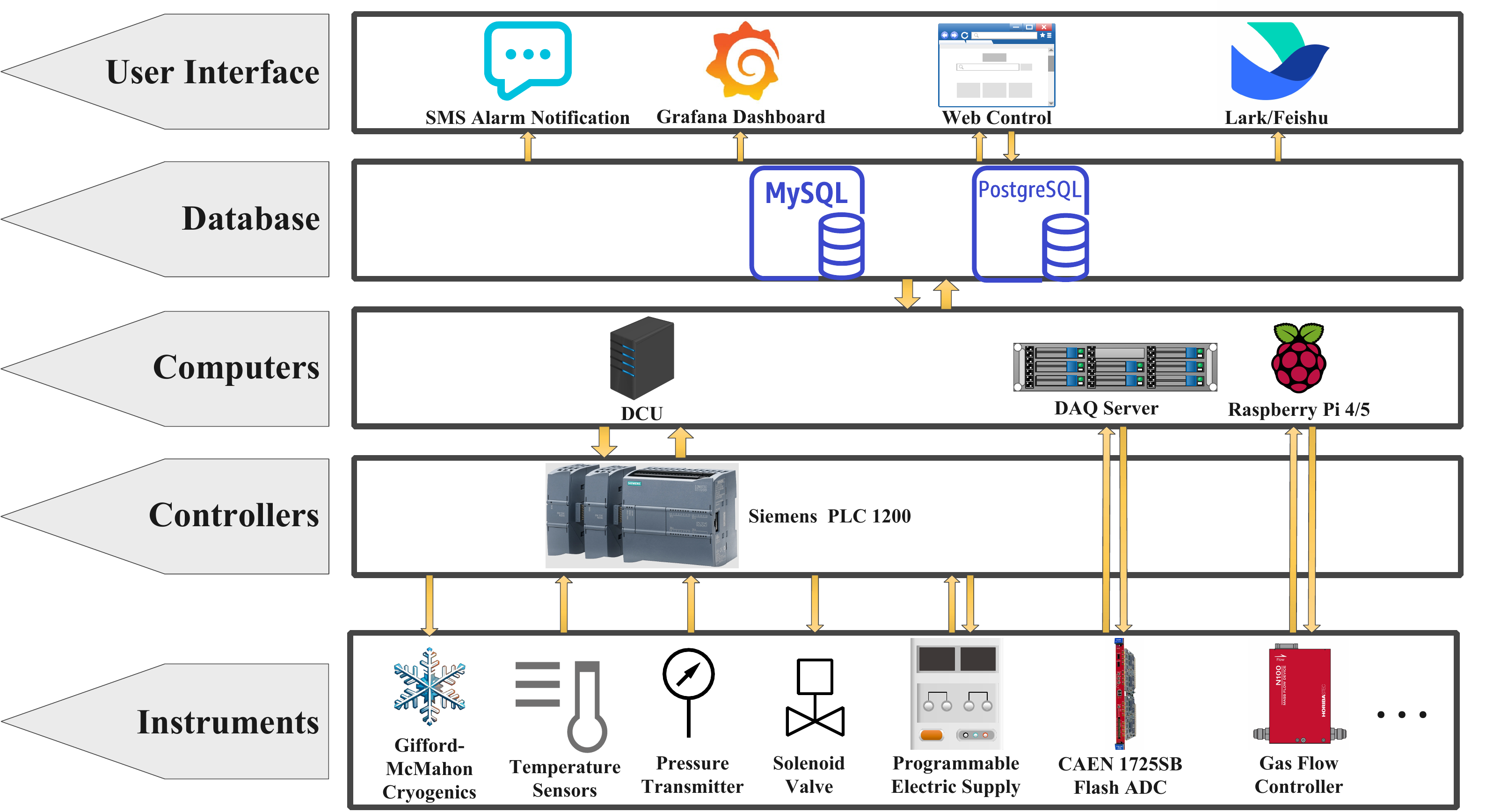}
    \caption{ Schematic of the RELICS semi-distributed SC architecture.
    The SC architecture includes the instruments, controllers, computers, database and user interface.
    Data flows upward for monitoring and the control commands propagate downward to instruments.}
    \label{fig:SC_system}
\end{figure*}

As shown in Fig.~\ref{fig:SC_system}, the system is centered around a \textit{Siemens S7-1200} programmable logic controller (PLC), which acts as the core processing unit to maintain stable detector performance. 
The PLC interfaces with a diverse array of sensors including silicon diodes, platinum resistance thermometers (PT100s), capacitive level meters, and pressure sensors to continuously monitor the detector state. 
It processes sampled data to log historical trends and issues control signals that autonomously regulate system parameters, such as heater power, through real-time feedback loops. 
Furthermore, the system incorporates robust safety features to prevent overpressure and overheating. 
Specifically, configurable alarm triggers can initiate automated emergency procedures, such as activating backup cooling or releasing xenon gas via a solenoid valve, thereby safeguarding the detector from potentially damaging conditions.

The Device Control Unit (DCU) is a personal computer responsible for the development and management of the detector control system. 
It is used to modify and edit the PLC program via Portal TIA, execute LabVIEW scripts to communicate with the PLC and upload data to the database, and control peripheral components such as the circulation pump and MFC through serial or other communication interfaces. 
All monitoring data and system logs are stored in the central database layer, implemented using MySQL and PostgreSQL.
The system employs Grafana as a graphical interface to monitor detector status and facilitate parameter adjustments.
In parallel, a dedicated user interface enables real-time monitoring and control of detector parameters, while a data acquisition system records signals from the detector sensors.
With scalability as a design goal, the DCU can support future expansion and integration with additional subsystems as needed.

\subsection{Flow rate control and monitoring}
\label{subsec:flow_control}

The overall circulation speed is driven by the bellows pump, whose motor is managed by VFD. 
This setup allows for broad control over the flow rate. 
The VFD is a controlling unit that communicates via the Mod-bus protocol, enabling remote monitoring of the pump frequency and power consumption. 
Furthermore, it is configured to issue an alarm and permit a remote restart in the event of an overload shutdown, enhancing the system operational robustness.

While the VFD controls the power, precise and quantitative control and measurement of the xenon flow rate are achieved using two high-precision MFCs. 
The flow rate is a critical parameter, as it directly impacts the efficiency of the purification process.
A higher flow rate generally corresponds to a faster rate of impurity removal. 
To this end, \textit{Horiba SEC-N100} MFCs, with a precision of $\mathrm{0.1\%}$, are installed in the primary circulation loop.  
These devices are controlled and monitored remotely via their serial port interfaces, providing the system with accurate, real-time flow rate data.

\subsection{Improved PID control for thermal regulation}
\label{subsec:PID}

Maintaining the thermal stability of the detector system at approximately $\mathrm{170~K}$ requires a reliable control system capable of balancing the constant cooling power of the GM cooler with a dynamically adjusted heat load. 
This is achieved using resistive heaters, whose current is governed by a feedback loop running on the PLC.
While a standard linear PID controller~\cite{lakeshore_pid_appendix} is effective for maintaining steady-state equilibrium, we found it inadequate for the highly dynamic phases of operation, such as xenon liquefaction, filling and setpoint adjustment. 
In these scenarios, the system experiences large, sudden thermal load variations. 
A standard linear controller typically suffers from slow response times to these massive disturbances or significant integral saturation, leading to severe overshoot during the stabilization phase. 
Even with signal filtering, the linear response does not effectively manage the non-linear thermodynamic behavior of the system during phase transitions.
We have therefore developed and implemented a custom, non-linear PID algorithm designed for enhanced robustness and adaptability~\cite{85c4d1c00a4b43a7a2aad5174c969473}.

The algorithm computes the heater current, $\mathrm{I_{heater}}$, from a nonlinear combination of proportional ($\mathrm{T_{P}}$), integral ($\mathrm{T_{I}}$), and derivative ($\mathrm{T_{D}}$) terms.
This formulation weights the dominant term more heavily while ensuring a damped overall response:

\begin{equation}
    \mathrm{ P_{heater} \propto I^2_{heater}(t)  = \sqrt{ \max\left( \sum_{\mathrm{j} \in \{\mathrm{P, I, D}\} } \mathrm{sgn}(\mathrm{T_j}) \cdot \mathrm{T_j^2}, ~0 \right) }}.
    \label{eq:pid_output}
\end{equation}

To ensure robust anti-saturation behavior and mitigate the effects of system fluctuations and sensor noise, a sign-preserving logarithmic function, $\mathrm{f(x) = \mathrm{sgn}(x) \cdot \log ( |x|+1 )}$, is applied to each term.
The three components are then calculated as:
\begin{align}
\mathrm{T_{P}} &= \mathrm{K_P \cdot f(\epsilon_p(t))} \\
\mathrm{T_{I}} & =\mathrm{K_I} \cdot \mathrm{f\left( \int_{0}^{t} \epsilon_p(\tau) d\tau \right)} \\
\mathrm{T_{D}} &= \mathrm{K_D \cdot f \left( \frac{d\tilde{p}(t)}{dt} \right)}
\end{align}
where $\mathrm{K_P, K_I, K_D}$ are the tunable gain coefficients. 
This root mapping was chosen empirically to provide a damped power response, which prevents heater power saturation during large disturbances while preserving fine control authority near equilibrium.
During stable operation, the same algorithm maintains pressure stability.
The relative pressure fluctuation, quantified as the standard deviation normalized to the setpoint, reaches $\mathrm{5.14\times10^{-4}}$, as summarized in Tab.~\ref{tab:alert_thresholds}.
Although the ultimate goal is thermal stability, the controller uses the pressure error $\mathrm{\epsilon_p(t) = p_{set} - \tilde{p}(t)}$ as its primary feedback, as pressure in a saturated system provides a faster and more sensitive indicator of the thermodynamic state than direct temperature measurements. 
Here, $\mathrm{\tilde{p}(t)}$ is obtained through a discrete-time Kalman filter that estimates the true pressure state from noisy measurements.

\subsection{Contingency handling and safety systems}
\label{subsec:contingency}

To ensure the safety of the detector, a multi-layered contingency handling system has been designed to autonomously manage off-nominal conditions. 
Such situations can arise from various scenarios, including power outages, loss of the primary water cooling circuit, or cryocooler performance degradation. 
The system is designed to provide robust protection against over-pressurization, which could otherwise lead to the loss of expensive xenon or damage sensitive components like the PMTs.

The first layer of protection is an automated alert system.
When any critical operational parameter deviates from its predefined safe range, the SC system immediately issues alerts. 
The specific set-points and emergency thresholds configured for the prototype operation are summarized in Tab.~\ref{tab:alert_thresholds}. 
When one of the parameters exceeds the upper limit, the SC system immediately issues alerts to laboratory personnel via designated communication channels, including E-mail, short message service, and instant messaging platform (Lark/Feishu).
This provides an early warning, allowing for timely manual intervention if necessary.

\begin{table}[htbp] 
\centering
\caption{Operational parameters and safety thresholds for the RELICS prototype.}
\label{tab:alert_thresholds}
\resizebox{\linewidth}{!}{
    \begin{tabular}{lcccc}
    \hline
    \textbf{Parameter} & \textbf{Setpoint} & \textbf{Fluctuation ($\sigma$)} & \textbf{Lower Limit} & \textbf{Upper Limit} \\
    \hline \hline
    Cryogenics pressure & 140 kPa & 0.072 kPa & 110 kPa & 170 kPa  \\
    Cryogenics temperature & --- & 0.153 K  & 150 K & 188 K \\
    Pump inlet pressure& --- & 0.155 kPa & 40 kPa & --- \\
    Pump outlet pressure & --- & 0.184 kPa & --- & 240 kPa \\
    \hline \hline
    \end{tabular}
}
\end{table} 

When the pressure continues to rise, the emergency cooling protocol is automatically triggered.
This involves activating a solenoid valve to introduce LN$_2$ into the pipes surrounding the cold head, providing additional cooling.
Due to thermal inertia and system response time, the cooling effect is not instantaneous.
The primary purpose of this LN$_2$ cooling is therefore to provide additional time for personnel to arrive on-site and address the root cause of the issue.

As a final and ultimate safeguard against rapid or extreme pressure excursions, two independent pressure relief mechanisms are in place. 
The first is an electronically controlled exhaust valve managed by the PLC, which will open if the pressure exceeds a critical software-defined threshold. 
To protect against a simultaneous power and control system failure, a purely mechanical pressure relief valve is also installed. 
This valve operates passively and is preset to open at $\mathrm{3~bar}$, preventing over-pressurization that could otherwise damage detector components, such as the PMTs.

\section{Data acquisition and processing System}
\label{sec:daq}

\subsection{Trigger and digital readout architecture}
\label{subsec:trigger}

Signals from the PMTs are directly digitized by a \textit{CAEN V1725SB} waveform digitizer. 
The \textit{V1725SB} is a 14-bit, 16-channel waveform digitizer featuring a sampling rate of $\mathrm{250~MS/s}$ and an analog bandwidth of 125~MHz, and an input dynamic range of $\mathrm{2~Vpp}$ per channel. 
To accommodate the high gain of the PMTs and preserve signal fidelity, no external amplification is utilized.

The PMT pulses are digitized at $\mathrm{4~ns}$ sampling intervals providing waveform reconstruction of the PMT signals.
The digitizer is configured using the \textit{CAEN A4818} driver, while data transmission to the acquisition server is managed by an \textit{A3818} PCIe controller through the \textit{CONET2} optical link protocol.
This system supports a maximum continuous data throughput of approximately $\mathrm{80~MB/s}$. 
When approaching this bandwidth limit, the digitizer may enter a “busy” state, potentially leading to event loss or dead time under high-rate conditions.

To enhance acquisition efficiency, the system supports the Digital Pulse Processing - Dynamic Acquisition Window (DPP-DAW) mode. 
In this mode, only waveform segments surrounding valid triggers are recorded, effectively suppressing zero-signal regions and reducing data volume. 
Each trigger pulse is shaped, integrated, and timestamped in real time, enabling efficient feature extraction for each event.

The digitizer supports two primary trigger modes:
\begin{itemize}
    \item \textbf{External Triggering}: Via TTL signal to the \texttt{TRG-IN} port, typically employed for LED PMTs gain calibration or synchronized coincidence measurements.
    \item \textbf{Self-Triggering}: Each channel autonomously generates a trigger when its input exceeds a predefined threshold. This mode is employed during standard TPC operation for detecting spontaneous events.
\end{itemize}

CAEN \texttt{DAW\_Demo} software is used to configure digitizer parameters and perform real-time waveform acquisition. 
Long-term performance monitoring, such as data rate, is conducted through Grafana dashboards.

The readout architecture is designed to be scalable. 
For the prototype detector, which employs 14~PMT channels, a single \textit{V1725SB} digitizer board provides sufficient readout capacity. 
In contrast, the full-scale RELICS detector will require the readout of approximately 150~PMT channels. 
To accommodate this increase, the system can be expanded by integrating multiple synchronized \textit{V1725SB} digitizer boards operating in parallel. 
Multi-board coordination is achieved through a common clock distribution and synchronized triggering network, ensuring seamless integration of the DAQ system within the main experimental infrastructure.

\subsection{Data processing and signal reconstruction}
\label{subsec:processing_framework}

The raw digitized waveforms are processed by a Python-based analysis framework based on the \texttt{strax} architecture~\cite{jelle_aalbers_2024_11355772}. 
This framework is designed for modularity and scalability, with its performance optimized using \texttt{NumPy} and \texttt{Numba}. 
All intermediate and final data products are stored in the compressed \texttt{NumPy} structured array format, with metadata described via JSON files.
The data processing pipeline comprises the following stages:
\begin{enumerate}
    \item \textbf{Binary}: Raw binary files generated by the DAQ system, containing unsorted and variable-length waveform segments.
    \item \textbf{Raw Records}: Sorted and vectorized raw waveform. These are reformatted into fixed-length chunks (e.g., 50 samples) and padded where necessary to ensure uniform size.
    \item \textbf{Records}: Baseline-subtracted and gain-calibrated waveforms. Baselines are computed using trimmed mean algorithms to reduce the influence of fluctuations or single-PE contamination. Channel-wise gain correction is applied to convert the waveform amplitude into the number of photoelectrons.
    \item \textbf{Peaks}: Candidate signal pulses reconstructed via software coincidence logic. Peaks from different channels are merged if their temporal separation is within a predefined coincidence window (typically 320~ns).
    \item \textbf{Events}: Physical events constructed by associating S1 and S2 peaks. Each event contains key observables including drift time, reconstructed horizontal position.
\end{enumerate}

Given that DPP-DAW mode records channel data independently, time-based software coincidence logic is essential for assembling true physical events. 
This strategy ensures that multi-channel signals are correctly identified as physical signals.

Overall, the processing framework is capable of efficiently transforming raw waveform data into physical datasets. 
Its modular design facilitates seamless adaptation to the full RELICS detector by extending processing resources and optimizing for higher data rates.
With minimal modification, this software pipeline can be deployed in the full experiment, supporting a fully paralleled analysis environment.

\subsection{Peak classification}
\label{subsec:s1s2_classification}

Building upon the established data processing framework, we classify the peaks recorded in the TPC based on their characteristics.
A series of threshold criteria and selections are applied to distinguish between different signal categories.
First, the pulse area is defined as the time integral of the baseline-subtracted waveform amplitude over the pulse window, which quantifies the total strength of the pulse. 
Second, the rise time characterizes the pulse shape and is defined as the interval between the $\mathrm{10\%}$ and $\mathrm{50\%}$ points of the cumulative area. 
Finally, the Area Fraction of Top (AFT) is defined as the fraction of the total area observed by the top PMT array:
\begin{equation}
  \mathrm{AFT =\frac{N_{{detected\; by\; top\; PMTs}}}{N_{{detected}}}}.
\end{equation}
  
\begin{figure}[!htbp]
    \includegraphics[width=\linewidth]{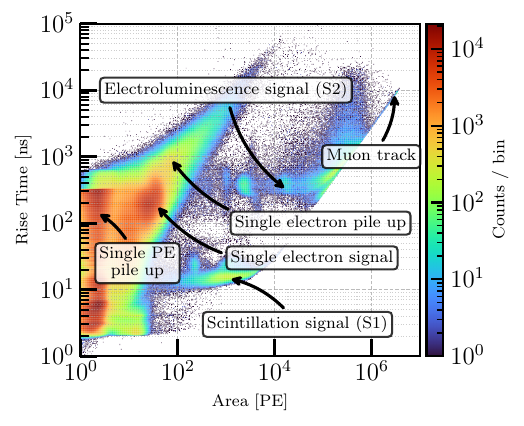}
    \caption{Distribution of rise time and area for all recorded peaks from a run with calibration source. This parameter distribution allows discrimination between different signal types. 
    Distinct populations including S1, S2, SE, calibration source and cosmic muon signals are visible.
    }
    \label{fig:peak_classification}
\end{figure}

As shown in Fig.~\ref{fig:peak_classification}, distinct signal clusters can be identified based on the distribution of peak area versus rise time:
\begin{itemize}
    \item \textbf{Scintillation signal (S1)}: Prompt, narrow pulses with fast rise times (typically $\mathrm{\sim 10~ns}$).
    \item \textbf{Electroluminescence signal (S2)}: Broader pulses with slower rise times (typically hundreds of ns).
    \item \textbf{Single electron signals}: Single electron (SE) signals are low-amplitude, S2-like pulses produced by the electroluminescence of an individual electron extracted into the gas phase~\cite{PhysRevD.106.022001}, which are analyzed in detail in Sec.~\ref{subsec:se_analysis}.
    \item \textbf{Single electron pile up}: The SE pile-up is low-charge S2-like signals appearing after large S2 events, which is the dominant background in low energy region.
    \item \textbf{Muon Track}: High amplitude ($\mathrm{Area~>~1\times~10^6~PE}$), wide ($\mathrm{Rise~time~>~1\times~10^4~ns}$) S2 signals caused by cosmic muons traversing the entire drift volume, often saturating the ADC and PMTs.
    \item \textbf{Single PE pile up}: Low area ($\mathrm{\sim 2~PE}$) signals below the detection threshold for classification, typically arising from noise, afterpulses, or uncorrelated photon events.
\end{itemize}
The peak classification plays a key role in separating background from signals. 
It also enables studies of temporal and spatial correlations between different signal classes, such as S1-S2 matching and DE recognition following high energy events.

\section{Detector data analysis}
\label{sec:analysis}

To evaluate the performance of the prototype TPC and establish a technical foundation for the calibration source preparation for full-scale RELICS experiment, we developed and deployed $\mathrm{^{37}Ar}$~\cite{guo2025preparationmeasurementrm37ar} and $\mathrm{^{83m}Kr}$~\cite{PhysRevC.105.014604} calibration sources via a dedicated injection system.
The following subsections detail the analysis of these peak- and event-level data, focusing on quantifying the light and charge yields as functions of electron recoil energy to validate the detector response.

\subsection{Single electron signal analysis}
\label{subsec:se_analysis}

SE signals represent the fundamental element of the S2 process and potentially arise from delayed extraction or spontaneous emission of electrons trapped at the liquid-gas interface.
They provide a sensitive benchmark for evaluating detector performance. 
The single-electron gain (SEG) is defined as the average integrated area of SE signals, typically expressed in units of photoelectrons per extracted electron (PE/e$^-$).
A higher SEG enhances the detector ability to resolve and count individual ionization electrons, thereby improving its sensitivity to low-energy interactions such as CE$\mathrm{\nu}$NS.
Accurate measurement of the SEG is therefore critical for reliable low-threshold event reconstruction.

\begin{figure}[!htbp] 
    \includegraphics[width=\linewidth]{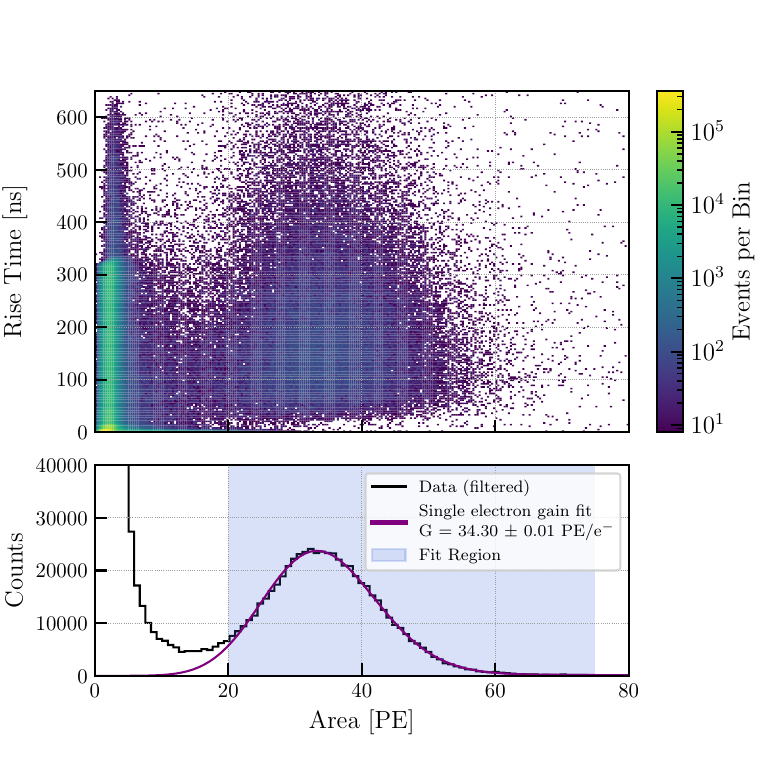}
    \caption{ \textbf{Top:} The two-dimensional distribution of S2-like signal area versus rise time for events satisfying the single electron selection criteria. 
    \textbf{Bottom:} The horizontal projection of the above distribution onto the area axis. The purple curve shows the best fit result of our single electron model to the data within the shaded fit region ($\mathrm{[20,75]~PE}$).
    }
    \label{fig:se_hist}
\end{figure}

To characterize the SEG, a population of SE pulses is selected.
The selection criteria include temporal isolation, requiring a window of $\mathrm{>100~\mu s}$ from preceding and subsequent S2 events, and a rise time filter $\mathrm{>40~ns}$  to reject S1 contamination~\cite{Aalbers:2018mfc}.
The resulting SE area distribution is presented in Fig.~\ref{fig:se_hist}.
The lower tail of the distribution is contaminated by low-amplitude backgrounds, primarily electronic noise and the pile-up of uncorrelated photoelectrons.
Consequently, to ensure a precise estimation of the SEG, the fit is applied exclusively within the range of $\mathrm{[20, 75]~PE}$, where the SE population is well-resolved and dominant.

To determine the SEG, we employ a mixture model that accounts for both the discrete statistics of PE detection and the contamination from double-electron peaks. 
The signal area $\mathrm{x}$ is modeled as a superposition of Gaussian distributions weighted by Poisson probabilities. 
First, we define the SE response function $\mathrm{\mathcal{S}(x; \lambda)}$ for an electron yielding a mean of $\lambda$ PEs as:

\begin{equation}
    \mathrm{\mathcal{S}(x; \lambda) = \sum_{n=1}^{\infty} \mathcal{P}(n; \lambda) \cdot \mathcal{N}(x; n, \sigma\sqrt{n})},
    \label{eq:response_func}
\end{equation}
where $\mathrm{\mathcal{P}(n; \lambda)}$ represents the Poisson probability of detecting $\mathrm{n}$ photoelectrons, and $\mathrm{\mathcal{N}(x; n, \sigma\sqrt{n})}$ describes the PMT resolution, modeled as a Gaussian centered at n with a width scaling as $\mathrm{\sqrt{n}}$.

The total fit function $\mathrm{f(x)}$ combines the contribution from SE events (with mean yield $\mathrm{G}$) and  two-electron events (with mean yield $\mathrm{2G}$):

\begin{equation}
\mathrm{
    f(x) = C_1 \cdot \mathcal{S}(x; G) + C_2 \cdot \mathcal{S}(x; 2G)}.
    \label{eq:se_mixture_model}
\end{equation}
Here, G represents the SEG (in $\mathrm{PE/e^-}$), while $\mathrm{C_1}$ and $\mathrm{C_2}$ are the normalization constants for the single- and double-electron components, respectively. 
By explicitly including the second term ($\mathrm{\mu = 2G}$), the model effectively decouples the influence of multi-electron pile-up, ensuring an unbiased estimation of the fundamental gain G.

Using this model, the fitted SEG is determined to be $\mathrm{(34.30 \pm 0.01~(stat.))~PE/e^-}$, as indicated in Fig.~\ref{fig:se_hist}, demonstrating stable and well-resolved SE signal behavior across the acquisition period.
The observed spectrum is well described by a convolution of a Poisson distribution (accounting for PE emission statistics) and a Gaussian function (modeling detector resolution).
In our prototype detector, the measured SEG average is approximately 34~PE/e$^-$, which meets the design requirement for resolving SEs and enables accurate reconstruction of small S2 signals in rare event searches.

\subsection{S2 X-Y position reconstruction}
\label{subsec:event_location}

\begin{figure}[!htbp]
    \includegraphics[width=\linewidth]{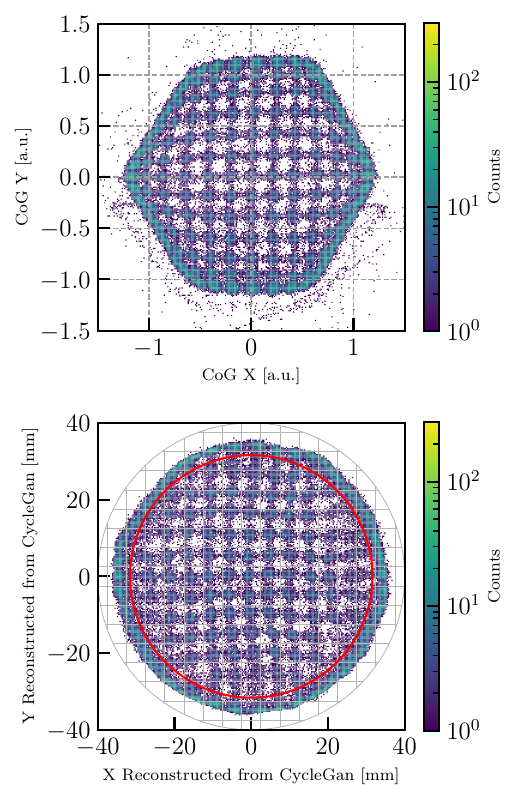}
    \caption{ Comparison of $\mathrm{^{83m}Kr}$ S2 event positions between CoG (top) and machine learning-based spatial reconstruction with domain adaptation (bottom). 
    Clusters correspond to electrons passing through the gate and being focused by the electric field. The red solid line represents the radius selection criteria.}
    \label{fig:position_reconstruction}
\end{figure}

Accurate event position reconstruction within the TPC is essential for background suppression and signal identification in the RELICS experiment.
The transverse $\mathrm{(x, y)}$ coordinates are inferred from the S2 light pattern on the top PMT array, while the $\mathrm{z}$ position is determined by the drift time between S1 and S2 signals~\cite{PhysRevD.111.062006}.
As shown in the top panel of Fig.~\ref{fig:position_reconstruction}, a simple Center-of-Gravity (CoG) calculation  yields only a rough position estimate,  leading to  spatial distortions due to geometry of the TPC.
Consequently, a Deep Residual Network (DeepResNet) combined with Cycle-consistent Generative Adversarial Network (CycleGAN)~\cite{guo2025domainadaptivepositionreconstruction} for domain adaptation is developed to enhance reconstruction accuracy.
The network takes pixelated S2 hit patterns as input and is trained on the Monte Carlo simulation mentioned in Sec.~\ref{subsec:simulation}.
Additionally, $\mathrm{^{37}Ar}$ K-shell decay calibration data are incorporated to compensate for discrepancies between the simulation and the actual detector response.

The performance of this method is shown in the bottom panel of Fig.~\ref{fig:position_reconstruction}.
The algorithm corrects the spatial distortions observed in the CoG reconstruction, mapping the events to their true physical coordinates.
As the localized spots caused by electrode grid are resolved, the position accuracy is estimated to be approximately $\mathrm{5~mm}$~\cite{Baudis:2020nwe}, which corresponds to the wire pitch.
Detailed implementation concerning the algorithm construction and quantitative error analysis are presented in Ref.~\cite{guo2025domainadaptivepositionreconstruction}.

\subsection{Simulation and calibration of the TPC light collection efficiency}
\label{subsec:simulation}

Particle interactions in the TPC produce S1 and S2 signals.
Both signals consist of VUV photons, generated in the liquid and gas phases, respectively.
A detailed understanding of the detector performance is crucial for both design and operational optimization.
For this target, a Geant4-based \textit{RelicsSim} software is developed~\cite{AGOSTINELLI2003250,  Chen_2021}. 
Within this framework, photons are treated as optical particles, meaning their propagation is governed solely by optical processes such as reflection, refraction, and absorption, while other particle physics interactions are disregarded.

The outcome of the optical simulation is quantified by the light collection efficiency (LCE), defined as a position-dependent quantity representing the percentage of photons emitted from a given position within the simulated detector volume that reach a given set of PMTs. It is calculated as:
\begin{equation}
\mathrm{
\text{LCE(x,y,z)} = \frac{N_{\rm{\gamma,detected}}(x,y,z)}{N_{\rm{\gamma,generated}}(x,y,z)}.
}
\end{equation}

For a given energy deposition, the amount of light measured by the PMT arrays depends on the interaction position within the TPC. 
This spatial dependence arises from factors such as VUV photon absorption in LXe, the reflectivity of detector materials, and the optical transparency of wire-mesh electrodes. 
To ensure the simulation accurately reproduces experimental data, the photon propagation model was specifically refined by tuning several key parameters in \textit{RelicsSim}:
\begin{itemize}
\item The VUV absorption length in LXe and GXe;
\item The reflectivity of PTFE surfaces in both LXe and GXe environments;
\item The refractive index of the PMT quartz windows.
\end{itemize}
Other optical parameters were fixed based on established literature values. 
The refractive index of xenon was based on previously published data~\cite{10.1063/1.2136879}, while the GXe layer between the LXe and the top PMT array was assigned a refractive index of 1. 
The Rayleigh scattering length was derived from theoretical calculations~\cite{Seidel:2001vf}. 
The unpolished PTFE surface was modeled with $\mathrm{99\%}$ diffuse reflectance, and the reflectivity of stainless steel was based on a prior measurement~\cite{Bricola:2007zz}.

To calibrate the detector response and validate our optical model, the prototype detector is injected with a $\mathrm{^{37}Ar}$ source.
$\mathrm{^{37}Ar}$ decays via electron capture to $^{37}$Cl with a half-life of 35.01 days, emitting a cascade of X-rays and Auger-Meitner electrons. This process produces distinct energy depositions corresponding to the K-, L-, and M-shell binding energies at $\mathrm{2.82~keV}$, $\mathrm{0.27~keV}$, and $\mathrm{0.01~keV}$, with branching ratios of $\mathrm{90.2\%}$, $\mathrm{8.7\%}$, and $\mathrm{1.1\%}$, respectively. For this analysis, the well-defined 2.82~keV K-shell line is selected to precisely calibrate the detector response.

A critical parameter for tuning the simulation is the AFT,  
which varies significantly with the light production position and is relevant to the optical parameters mentioned earlier. 
Importantly, it eliminates dependence on the unknown initial number of generated photons in the data, making it an ideal observable for model tuning. 
The tuning procedure involved an iterative comparison, where we simulated events distributed uniformly throughout the TPC volume and calculated the AFT as a function of depth. 
The key optical parameters were then scanned across appropriate ranges until the simulated depth-dependent S1 AFT distribution and the S2 AFT value achieved the best possible match with the one measured from the $\mathrm{^{37}Ar}$ calibration data.
The resulting parameters are summarized in Tab.~\ref{table:para}.

\begin{table}[htp]
	\begin{center}
		\caption{ Optical model parameters used in \textit{RelicsSim}, optimized to reproduce the experimental data.}
		\label{table:para}

			\begin{tabular}{ccc} \hline \hline 
				 Quantity  &Liquid &Gas    \\  \hline
				Absorption length (m)  & 5 & 50\\  
				PTFE reflectivity (\%) & 99 &70\\ 
				Refractive index of the PMT quartz window  & 1.7 &1.7\\  \hline \hline 	
			\end{tabular}
		
	\end{center}
\end{table}

\begin{figure}
\centering
\includegraphics[width=\linewidth]{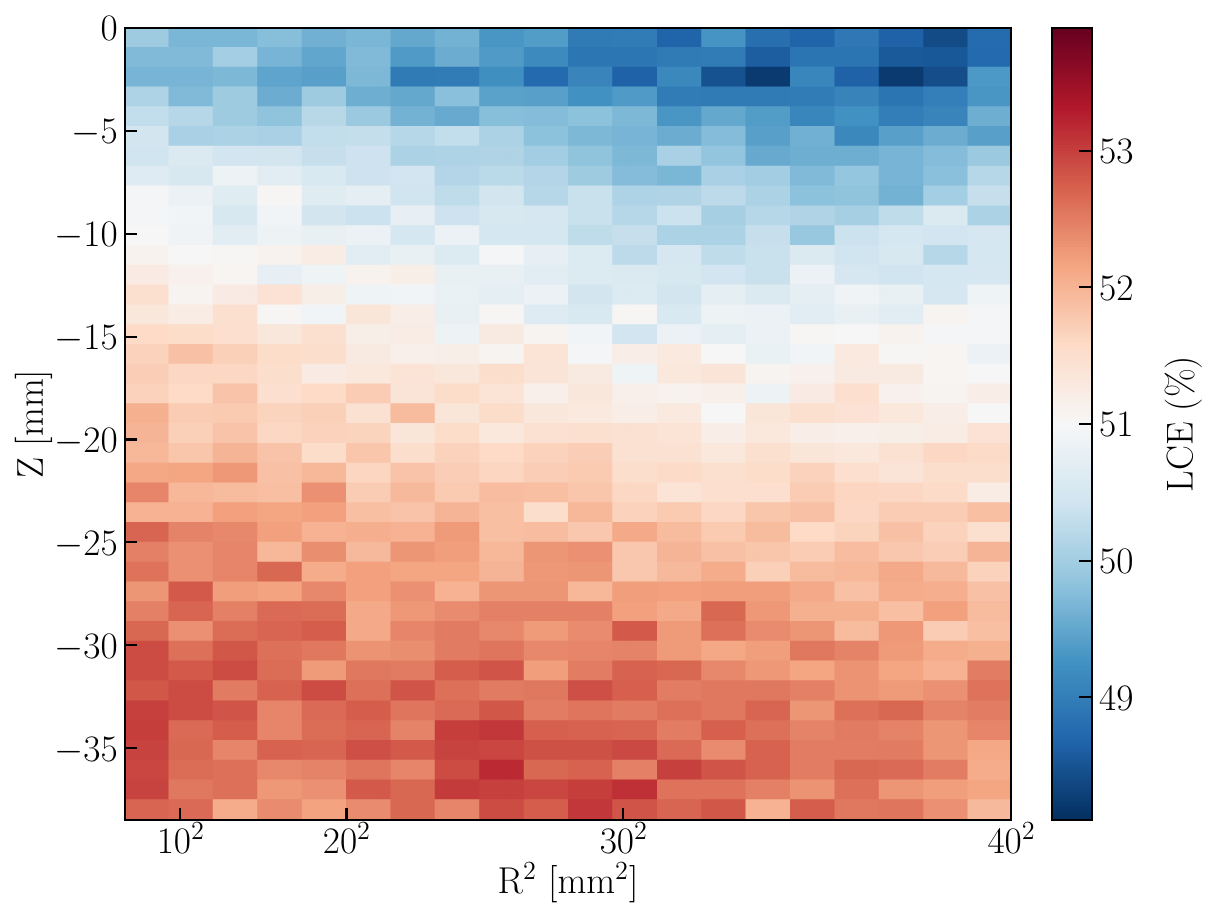}
\caption{Simulated LCE map for S1 signals. Z = 0 corresponds to the position of the LXe surface. The map generated using the tuned optical parameters reveals that the LCE is highest near the bottom-center of the TPC and decreases towards the top and outer radius.}
\label{drift_region_LCE}
\end{figure}

For the S1 signal, photons are generated iso-tropically and uniformly within the liquid drift region. 
The resulting axially symmetric S1 LCE map is presented as a function of vertical position Z and squared radius R$^2$ in Fig.~\ref{drift_region_LCE}. 
The volume-averaged S1 LCE, $\mathrm{\overline{LCE}_{S1,sim}}$, is determined to be $\mathrm{51.3\%}$.

\begin{figure}
	\centering
	\includegraphics[width=1\linewidth]{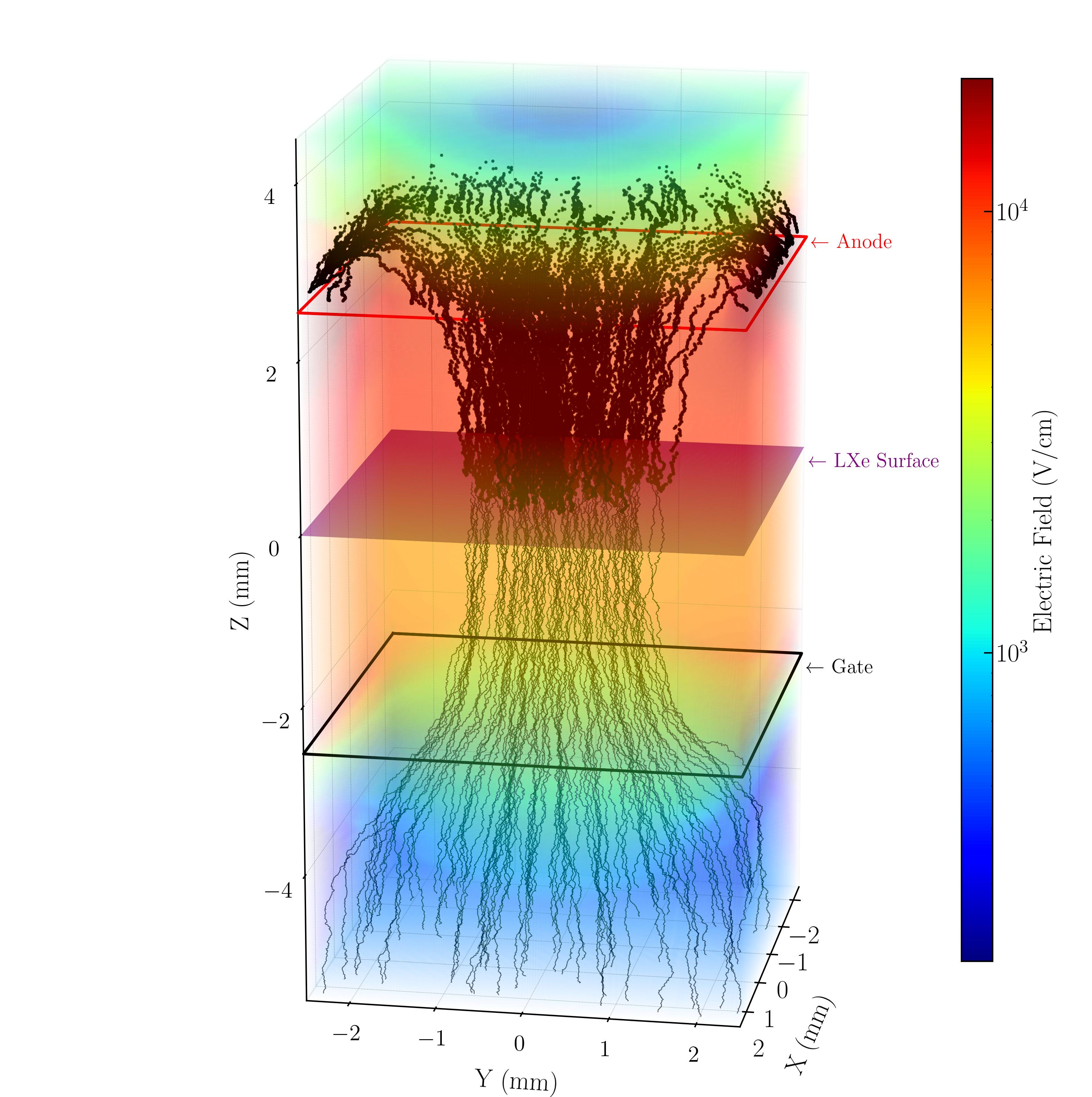}
	\caption{Electroluminescence simulation in a single grid cell using Garfield++.
    The black lines represent the drift trajectories of individual electrons originating in the LXe, and the black points indicate photon emission sites generated by electrons accelerating in the high-field gas region.
    The background color visualizes the magnitude of the electric field, distinguishing the weaker drift field in the liquid from the very strong amplification field in the gas.}
	\label{S2_sim}
\end{figure}

The generation of the S2 signal via proportional electroluminescence in the gas gap was modeled using the Garfield++~\cite{Garfieldpp} coupled optical simulation shown in Fig.~\ref{S2_sim}. 
Based on this simulation, the volume-averaged S2 LCE, $\mathrm{\overline{LCE}_{S2,sim}}$, is determined to be $\mathrm{48.5\%}$.

In summary, the \textit{RelicsSim} simulation, tuned via AFT matching against $\mathrm{^{37}Ar}$ calibration data, determined the key optical model parameters. 
Consequently, the S1 LCE map derived from this model serves as a critical basis for signal correction in the subsequent analysis. 
This capability not only validates our understanding of the true optical parameters but also provides a reliable model for the future RELICS experiment, laying the foundation for an optimized detector design.

\subsection{Event reconstruction and correction}
\label{sec:event_reco_and_corr}

To accurately determine the energy of an interaction within the dual-phase xenon TPC, the raw detector signals, S1 and S2, must be corrected for various position-dependent effects. 
The procedure yields corrected signals, denoted cS1 and cS2, which are independent of the event position and directly proportional to the initial number of photons and electrons produced.
This process involves two main stages: precise 3D position reconstruction followed by position-based signal correction.

\subsubsection{3D position reconstruction}

The reconstruction of the initial 3D vertex (x$_0$, y$_0$, z$_0$) of an event is essential for the analysis.
This is achieved through a multi-step process that combines the effect of electric field, electron drift lifetime, light collection efficiency.

First, as the horizontal position (x$\rm_r$, y$\rm _r$) is reconstructed from the S2 signal hit pattern observed on the top PMT array, as discussed in subSec.~\ref{subsec:event_location}. 
With the reconstructed horizontal position (x$\rm_r$, y$\rm _r$) and the measured electron drift time $\mathrm{t_d}$, the initial vertical position $\mathrm{x_0}$,  $\mathrm{y_0}$,  $\mathrm{z_0}$ is determined by tracing the electron cloud path back from the liquid-gas interface to its origin. 
This requires a precise model of the electron drift velocity field $\mathrm{\vec{v}(x, y, z)}$ inside the drift region. 
The back-tracing itself is performed by numerically solving the equation of motion $\mathrm{d\vec{r}/dt = -\vec{v}(\vec{r})}$ backward in time. 
We use the Runge-Kutta method starting from the endpoint (x$_{\mathrm{r}}$, y$_\mathrm{r}$, z$_{\text{gate}}$) and integrating over the duration $\mathrm{t_d}$ to find the initial position (x$_0$, y$_0$, z$_0$).

\subsubsection{Electron lifetime estimation}
\label{subsec:lifetime}

The ability to drift ionization electrons over distances on the order of tens of millimeters with minimal loss is a fundamental requirement for the performance of a dual-phase xenon TPC. 
The primary mechanism for electron loss during drift is attachment to electronegative impurities, such as O$_2$ and H$_2$O, that may be present in the xenon or emitted from detector materials. 
A key parameter quantifying xenon purity is the electron lifetime, $\mathrm{\tau_e}$, defined as the mean time that ionized electrons can drift in xenon before being captured by impurity molecules. 
The surviving fraction of electrons, $\eta$, after a drift time $\mathrm{t_d}$ follows an exponential attenuation law:
\begin{equation}
    \label{eq:eta}
    \mathrm{\eta(t_d) = \exp(-t_d / \tau_e)}.
\end{equation}

To measure $\mathrm{\tau_e}$ in the RELICS prototype, we use the $\mathrm{41.5~keV}$ total energy deposition from the $\mathrm{^{83m}Kr}$ cascade decay, reconstructed by summing the signals from the $\mathrm{32.1~keV}$ and $\mathrm{9.4~keV}$ transitions.
This source is injected as a gaseous calibration source and uniformly distributed in the active volume. 
Events within a central fiducial radius ($\mathrm{r^2 < 1000~mm^2}$) are selected to ensure uniform detector response, while regions near the gate and cathode (drift time $<5$~\textmu s or $>20$~\textmu s) are excluded to avoid electric field non-uniformities.
Since the initial ionization yield is constant for these events, the decrease in the position-corrected S2 signal can be attributed to electron attachment. 
In this analysis, we use an S2 signal corrected for spatial non-uniformity only, denoted as $\mathrm{scS2}$, to isolate the loss effect along the drift time.
The data are binned by drift time, and the mean $\mathrm{scS2}$ value is computed for each bin at a certain drift time.

\begin{figure}[!htbp]
	\centering
	\includegraphics[width=\linewidth]{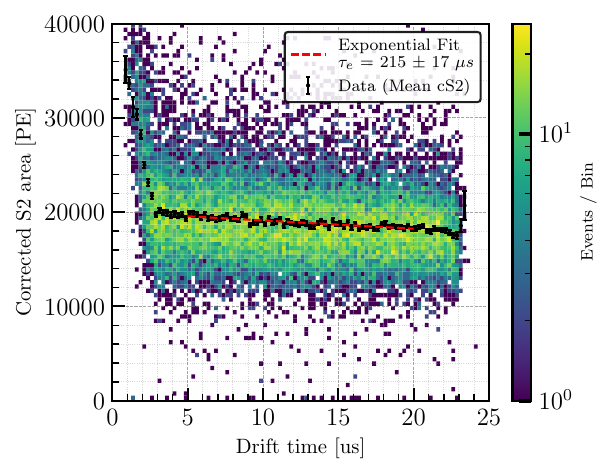}
    \caption{
    Measurement of the electron lifetime using total $\mathrm{41.5~keV}$ $\mathrm{^{83m}Kr}$ decay events.
    The 2D histogram shows the spatially corrected S2 area ($\mathrm{scS2}$) as a function of electron drift time for events within a central fiducial volume.
    The black data points indicate the mean $\mathrm{scS2}$ in each drift time bin, fitted with an exponential decay (red dashed line).
    The fit yields an electron lifetime of $\mathrm{215 \pm 17~\mu s}$.
   }
    \label{fig:electron_lifetime}
\end{figure}

Figure~\ref{fig:electron_lifetime} shows the dependence of $\mathrm{scS2}$ on drift time, where the exponential decrease signifies charge attenuation due to electronegative impurities.
The fitted lifetime of $\mathrm{\tau_e = 215 \pm 17~\mu s}$ indicates the xenon purity level in the detector.
This result was obtained at a circulation rate of $\mathrm{12.78~SLPM}$.

The electron lifetime in this prototype was primarily limited by the use of large PTFE fillers inside the cryostat, which were installed to displace volume and minimize xenon consumption.
These fillers introduced a significant surface area and trapped impurities, leading to a high outgassing load. 
Furthermore, the measurement of electron lifetime is constrained by the short drift length and non-uniformities in the electric field.
For the full-scale RELICS experiment, a millisecond-scale lifetime is required. 
The purification capacity will be enhanced through an optimized gas handling system with reduced flow impedance and higher-performance bellows pumps, enabling a higher circulation flow rate around $\mathrm{100~SLPM}$.

\subsubsection{Signal correction}
\begin{figure}[!htbp]
	\centering
	\includegraphics[width=\linewidth]{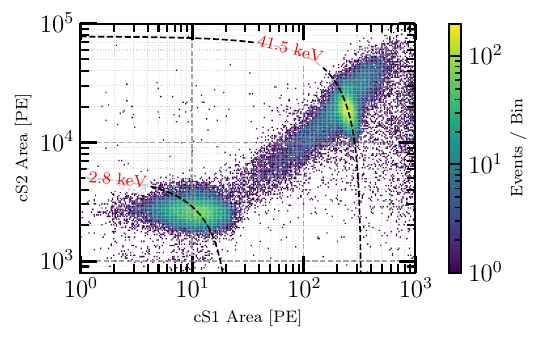}
    \caption{ Two-dimensional distribution of corrected S2 versus corrected S1 of calibration data using $\mathrm{^{37}Ar}$ and $\mathrm{^{83m}Kr}$ sources. The main panel shows the logarithmic event density with distinct populations from the 2.8~keV K-shell electron capture of $\mathrm{^{37}Ar}$ (lower left) and the 41.5~keV cascade decay of $\mathrm{^{83m}Kr}$ (upper right). The black dashed curves represent theoretical iso-energy contours calculated from the detector calibrated light yield and charge yield.}
    \label{fig:s2_vs_s1_calib}
\end{figure}

Once the 3D position and drift time of events are known, the S1 and S2 signals can be corrected for their spatial dependencies. 
The S2 signal is affected by spatial non-uniformities in the drift field and charge loss due to electron attachment to impurities.
The S1 signal exhibits a spatial dependence due to two primary effects: the position-dependent LCE, governed by geometry and reflections, and the field-dependent L$_\mathrm{y}$, which varies with the local electric field due to electron-ion recombination effects.
The distribution of the corrected event-level signals is illustrated in Fig.~\ref{fig:s2_vs_s1_calib}.

The corrected S2, denoted as cS2, is obtained by:
\begin{equation}
    \mathrm{cS2 = S2_{\text{raw}} \cdot \exp\left(\frac{t_d}{\tau_e}\right) \cdot \frac{Q_{y, 0}}{Q_y(E(x, y, z))}},
\end{equation}
where the exponential term accounts for the attenuation based on the electron lifetime $\mathrm{\tau_e}$ and the drift time $\mathrm{t_d}$. 
The second term corrects for the field-dependent charge yield $\mathrm{Q_y}$, where the local electric field $\mathrm{E}$ at the interaction vertex $\mathrm{(x, y, z)}$ is derived from simulations, and the corresponding yield is calculated using the NEST model. 
$\mathrm{Q_{y, 0}}$ denotes the charge yield at the geometric center of the drift region, ensuring that cS2 is normalized to the response at the detector center.

Similarly, the S1 area is corrected for the position-dependent LCE and the field-dependent light yield $\mathrm{L_y}$.
The corrected S1 signal is defined as:

\begin{equation}
    \mathrm{cS1 = S1_{\text{raw}} \cdot \frac{\text{LCE}_0}{\text{LCE}(x, y, z)} \cdot \frac{L_{y,0}}{L_y(E(x, y, z))}},
\end{equation}

where $\mathrm{LCE_0}$ and $\mathrm{L_{y,0}}$ denote the LCE and $\mathrm{L_y}$ at the geometric center of the TPC.
This normalization ensures that cS1 represents the signal response as if the interaction occurred at the detector center.

\subsubsection{Energy calibration and resolution}
\label{sec:energy_calib}

To calibrate the response of the prototype TPC, the primary scintillation gain $\mathrm{g_1}$ and electroluminescence gain $\mathrm{g_2}$ are estimated.
Events from $\mathrm{^{37}Ar}$ K-shell decay and $\mathrm{^{83m}Kr}$ decay are utilized for calibrating.
The $\mathrm{^{37}Ar}$ source provides a $\mathrm{2.82~keV}$ peak from K-shell electron capture.
For the $\mathrm{^{83m}Kr}$ source, we utilize the total energy deposition of $\mathrm{41.5~keV}$, following the same procedure as described in subSec.~\ref{subsec:lifetime}.

\begin{figure}[!htbp]
    \centering
    \includegraphics[width=\linewidth]{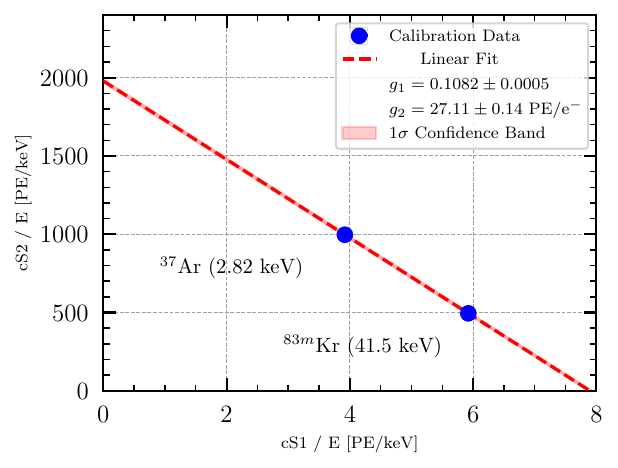}
    \caption{The energy-normalized S2 signal versus the S1 signal for calibration data from $\mathrm{^{37}Ar}$ (2.82 keV) and $\mathrm{^{83m}Kr}$ (41.5 keV). The data points are fitted with a linear function (red dashed line) to extract the primary scintillation gain ($\mathrm{g_1}$) and the charge gain ($\mathrm{g_2}$). The 1$\sigma$ confidence band of the fit is shown as the shaded red area.}
    \label{fig:energy_calib}
\end{figure}

The deposition energy in LXe is partitioned into scintillation photons (S1 signal) and ionized electrons (S2 signal). 
This partitioning exhibits an anti-correlation for electron recoil events. 
The relationship between the corrected signals (cS1 and cS2) and the energy E can be expressed as:
\begin{equation}
    \label{eq:energy_calib}
    \mathrm {E = W \left( \frac{\text{cS1}}{g_1} + \frac{\text{cS2}}{g_2} \right),}
\end{equation}
where $\mathrm{W = 13.7~eV}$ is the average work function to produce a quantum (either a photon or an electron) in LXe. 
Figure~\ref{fig:energy_calib} illustrates the result of this calibration procedure. 
The mean values of the cS1/E and cS2/E distributions for the two calibration peaks are plotted. 
A linear fit to these points allows for a direct extraction of the gain factors. 
The fit yields a primary scintillation gain of:
\begin{equation}
\label{g1}
\mathrm {g_1 = 0.1082 \pm 0.0005~PE/ph} 
\end{equation}
and the electroluminescence gain of:
\begin{equation}
\label{g2}
 \mathrm{g_2 = 27.11 \pm 0.14~PE/e^-} 
\end{equation}

Since $\mathrm{g_2}$ represents the gain per ionized electron, while the SEG corresponds to the gain per extracted electron, the two quantities differ by the EEE.
The ratio $\mathrm{g_2/SEG = (79.04 \pm 0.41) \%}$ falls in the range of $70\%$ to $90\%$ as estimated in Sec.~\ref{subsec:field_cage}.
We also estimated the $\mathrm{g_1}$ based on the optical simulation. This value, denoted as $\mathrm{\bar{g}_{1,sim}}$, is defined as:
\begin{equation}
    \mathrm{\bar{g}_{1,sim} = \overline{LCE}_{S1,sim} \cdot QE\cdot CE}.
\end{equation}
Among the PMTs used in the prototype, the quantum efficiency (QE) and the collection efficiency (CE) from the photocathode to the first dynode are $\mathrm{(33\pm2)\%}$ and $\mathrm{(75\pm7)\%}$, respectively~\cite{hamamatsu_private}. The simulation model predicts the $\rm \bar{g}_{1,sim}$ of $(0.127\pm 0.015)$~PE/ph, which is consistent with Eq.~\ref{g1} within 2$\sigma$.
Similarly, the predicted electroluminescence gain, $\mathrm{\bar{g}_{2,sim}}$ is calculated as 
\begin{equation}
\mathrm{\bar{g}_{2,sim} = N_{\gamma}\cdot \overline{LCE}_{S2,sim} \cdot QE\cdot CE\cdot EEE}.
\end{equation}
Here, $\mathrm{N_{\gamma} = 288}$ is the average photon yield per electron obtained from the Garfield++ simulation at the prototype extraction field.
The simulation model predicts a $\rm \bar{g}_{2,sim}$ of $(24.35\pm 2.69)$ PE/e$^-$, which is consistent with Eq.~\ref{g2}.

The detector energy resolution, defined as the ratio of the standard deviation to the peak energy, is measured at the two calibration points and fitted using the empirical relation: 
\begin{equation}
\label{eq:energy_res}
\mathrm {\frac{\sigma(E)}{E} = \frac{30.6 \%}{\sqrt{E~(keV)}} + 2.6\%},
\end{equation}

where E is the deposited energy in keV. 
This formula offers a simple description of the detector energy resolution from 2.8 keV to high 41.5 keV energies.

\subsection{S2-only analysis demonstration and background rate estimation}
\label{subsec:s2_only}

\begin{figure}[!htbp]
\includegraphics[width=\linewidth]{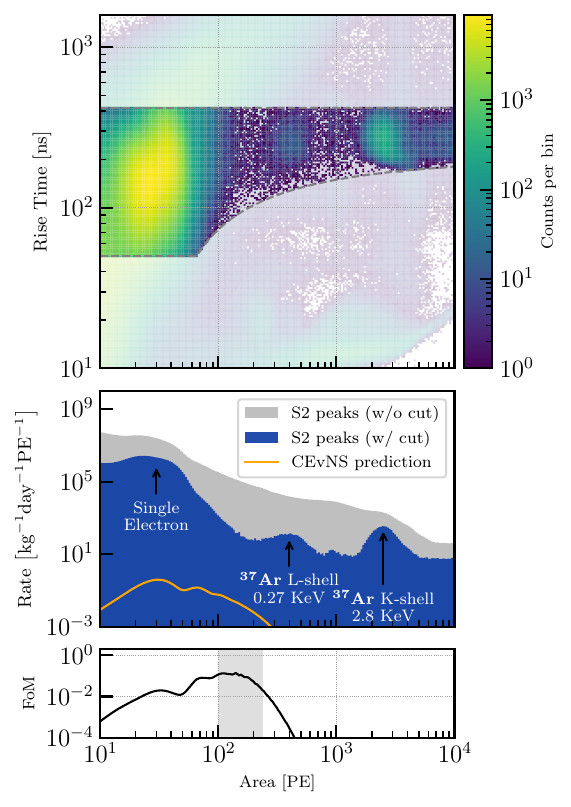}
\caption{
Demonstration of the S2-only analysis strategy achieving sub-keV sensitivity.
\textbf{Top:} Two-dimensional distribution of S2 signal area (PE) versus rise time (ns).
The comparison between all detected peaks (shadowed gray) and the selected S2 signals, after applying the high energy veto, fiducial volume selection, and rise time cut, illustrates the reduction of the DE background. 
The dense cluster at S2 signals of several hundred PE and long rise times corresponds to high energy events producing DEs.
\textbf{Middle:} 
Event rate as a function of S2 area.
Before cuts (gray), the spectrum is dominated by DEs, and the $\mathrm{^{37}Ar}$ L-shell peak is obscured.
After applying high energy veto, fiducial volume selection, and rise time cut (blue), distinct $\mathrm{^{37}Ar}$ L-shell (0.27~keV) and K-shell (2.8~keV) peaks emerge, confirming sensitivity to sub-keV energy depositions and validating the effectiveness of the background rejection strategy.
\textbf{Bottom:}
Estimated significance ($\mathrm{N_{sig}/\sqrt{N_{bkg}}}$) as a function of area, calculated for an exposure of $\mathrm{32~kg \times 365~days}$. 
The shaded gray region marks the region of interest (ROI) for CEvNS searches, corresponding to signals between 100 and 240~PE.
}
\label{fig:s2_only}
\end{figure}

The CE$\mathrm{\nu}$NS search analysis relies on amplified ionization signals, since the primary scintillation from sub-keV nuclear recoils is too weak to be observed.
The region of interest (ROI) for CE$\mathrm{\nu}$NS is the S2 signal range of $\mathrm{ \left[ 100, 240 \right] }$~PE.
To assess the detector sensitivity to sub-keV interactions and quantify residual backgrounds, an S2-only analysis was conducted, as illustrated in Fig.~\ref{fig:s2_only}.

The dominant background in this regime originates from DE emission following high energy events.
These DEs produce S2 signals whose areas overlap with those of genuine sub-keV interactions, and thus constitute the dominant instrumental background in CE$\mathrm{\nu}$NS searches.
To suppress such events, a high energy veto is implemented: any S2 candidate occurring within a time window $\mathrm{T_{veto}}$ after a preceding large S2 pulse ($>$10$^5$~PE) is rejected,
\begin{equation}
\mathrm{T_{veto} = (20 \times Area_{HE} + 1.7\times10^6)~ns.}
\label{eq:HE_cut}
\end{equation}
This criterion suppresses DE afterpulses while preserving genuine low-energy signals.
A subsequent fiducial volume cut ($\mathrm{R^2 < 10~cm^2}$) further reduces edge-related backgrounds.

Without the high energy veto, the intense DE background obscures the $\mathrm{^{37}Ar}$ L-shell peak in the raw S2 spectrum.
After all cuts, distinct peaks from internal calibration sources ($\mathrm{^{37}Ar}$ L-shell at 0.27~keV and K-shell at 2.8~keV) can be clearly resolved, which confirms sub-keV detection capability with S2-only analysis.

The total S2 event rate before selection is $\mathrm{3.77\times10^{7}}$~events/(kg$\cdot$day) within the CE$\mathrm{\nu}$NS ROI.
After applying the high energy veto, rise time and fiducial volume cuts, the residual event rate is reduced to $\mathrm{8.57\times10^{3}}$~events/(kg$\cdot$day), which remains approximately four orders of magnitude higher than the expected CE$\mathrm{\nu}$NS signal rate of 0.98~events/(kg$\cdot$day). 
The uncertainty of the fiducial volume cut is estimated to be $\mathrm{31.6\%}$ through propagating position reconstruction uncertainty, as described in subSec.~\ref{subsec:event_location}.
This uncertainty does not affect the order-of-magnitude estimation of the signal and background event rates.

The DE background rate in the prototype remains four orders of magnitude higher than the expected CE$\mathrm{\nu}$NS signal rate. 
This excess arises primarily from multiple experimental limitations intrinsic to the small scale prototype setup.
First, high energy S2 events induced by external $\gamma$-rays and cosmic muons are the dominant source of DEs.
In the prototype, the absence of passive shielding results in a $\gamma$-rays interaction rate of order $\mathcal{O}(100)\mathrm{Hz}$, which is approximately two orders of magnitude higher than full-scale RELICS detector. Furthermore, the muon rate is $\mathcal{O}(1)\mathrm{Hz}$ and a factor of 10 lower than the full-scale RELICS detector.
Consequently, the average time interval between high energy interactions is reduced to approximately $\mathrm{10~ms}$, compared to about $\mathrm{100~ms}$ expected in RELICS.
Since the DE rate follows a power-law dependence on the time gap following high energy events, this shorter interval contributes to an enhancement of the DE background by more than an order of magnitude.
Second, the limited number of PMTs constrains position reconstruction, preventing effective space-time correlation vetoes against DE originating near a previous high energy event.
In addition, the anode readout suffers from signal saturation under muon or high energy S2 signals.
The PMT saturation effect impedes identification and tracking of muon trajectories, which are essential for subsequent DE rejection.
These are sources of one order of magnitude background suppression.
The full-scale RELICS detector will adopt a dual-readout configuration, adding signals from an intermediate PMT dynode to the standard anode readout. 
This strategy extends the dynamic range, ensuring that muon tracks and high energy events are precisely characterized without saturation~\cite{Yang_2026}.
Moreover, 3D reconstruction of the muon track will be performed to suppress the DE background.
Finally, DE backgrounds often comprise the superposition of electrons emitted from spatially distributed positions associated with prior muon tracks and can be distinguished via analyzing their waveforms and distribution patterns on the top PMT array. This pattern analysis is expected to provide an order of magnitude rejection power in the full-scale detector.
In conclusion, the full-scale RELICS detector will overcome these limitations through enhanced position reconstruction enabled by a larger PMT array, PMT dual-readout, passive shielding, and space-time correlation algorithms~\cite{cai2024relicsreactorneutrinoliquid}, which together are expected to suppress DE backgrounds by approximately three orders of magnitude relative to the prototype to reach the target sensitivity.

Integrating the signal and background rates over the CE$\mathrm{\nu}$NS ROI from 100 to 240 PE, we obtain a total projected signal significance of $\mathrm{0.52\sigma}$ for an assumed exposure of $\mathrm{32~kg \times 365~days}$. 
Reducing background approximately two orders of magnitude relative to the current prototype can achieve a $5\sigma$ significance within this ROI, which sets a quantitative requirement for the background suppression to be achieved by the dual-readout, passive shielding, and space–time veto in the full RELICS detector.

\section{Conclusion}
\label{sec:conclusion}

The potential of the RELICS experiment in detecting CE$\mathrm{\nu}$NS depends on achieving a low energy detection threshold, which requires stable detector operation, high electron and photon collection efficiencies, and robust data analysis capability.
The development and operation of the dual-phase xenon TPC prototype have established the essential technical foundation for the RELICS experiment.
All detector subsystems including the time projection chamber, cryogenics and purification systems, slow control, and data acquisition have been designed, integrated, and validated under realistic operational conditions.
The validated instruments and technologies will be directly implemented in the construction and operation of the full-scale RELICS detector.

In parallel, an analysis and simulation framework has been developed to support event reconstruction, detector modeling, and signal correction, which will also be applied to the formal experiment.
The prototype has experimentally demonstrated the feasibility of a dual-phase xenon detector for low-energy reactor antineutrino studies, achieving a high SEG of $\mathrm{(34.30 \pm 0.01 (stat.))~PE/e^-}$, a calibrated charge gain $\mathrm{g_2}$ of $\mathrm{(27.11 \pm 0.14)~PE/e^-}$ and successfully detecting 0.27~keV L-shell decay events from $\mathrm{^{37}Ar}$.

Further improvements have been identified, including optimization of electrode biasing to enhance the SEG, surface polishing of PTFE reflectors to increase light collection efficiency, and refined field-shaping ring geometry to improve electric field uniformity.

In summary, the RELICS prototype has demonstrated the viability of a dual-phase xenon detector optimized for sub-keV energy sensitivity, providing a technological and analytical basis for the forthcoming full-scale RELICS experiment.

\vspace*{5ex}      
\noindent  \textbf{Acknowledgements}
RELICS is supported by grants from National Key R\&D program from the Ministry of Science and Technology of China (No. 2021YFA1601600), Natural Science Foundation of China (Nos. 12521007, 12275267, 12405129, 12375095, 12250011, 12305121), Ministry of Education of China (No. SRICSPYF-ZY2025028), Beijing Natural Science Foundation (Nos. QY23088, QY25008), Guangzhou Municipal Science and Technology Project (No. 2025A04J5409), Zhejiang Provincial Natural Science Foundation of China (No. LQKWL26A0501), and the CUHK-Shenzhen University Development Fund (No. UDF01003491).
We acknowledge CNNC Sanmen Nuclear Power Company for hosting RELICS.

\bibliographystyle{spphys2}

\bibliography{references}

\end{sloppypar}
\end{document}